\documentclass[12pt]{article}

\usepackage[a4paper,margin=1in]{geometry}
\usepackage[T1]{fontenc}
\usepackage[utf8]{inputenc}
\usepackage{lmodern}

\usepackage{graphicx}
\usepackage{xcolor}
\usepackage{amsmath,amssymb,amsthm}
\usepackage[numbers,sort&compress]{natbib}
\usepackage[colorlinks=true,linkcolor=blue,citecolor=blue,urlcolor=blue]{hyperref}
\usepackage{cleveref}
\usepackage{setspace}
\usepackage{lineno}

\graphicspath{{na35_plots/}{na49_plots/}{na61_plots/}}

\title{Heavy-ion physics at the CERN SPS H2: \\
NA35, NA49 and NA61/SHINE \\[2mm]
\Large with personal recollections}

\author{Marek Gazdzicki\\
\small Jan Kochanowski University, Kielce, Poland\\
\small \texttt{Marek.Gazdzicki@cern.ch}}

\date{}

\begin{document}

\maketitle

\begin{abstract}
This review presents a unified account of the NA35, NA49, and NA61/SHINE experiments, which together form a continuous programme of heavy-ion studies conducted at the H2 beamline of the CERN North Area using the SPS accelerator.
The programme, spanning about 40 years, was driven by the search for a high-density state of strongly interacting matter---the quark--gluon plasma (QGP)---and the transitions leading to it. The review focuses on this primary line of research.
The highlights of the programme include the observation of the first signal of QGP creation at the top SPS energy in S+S collisions by NA35, evidence for the onset of deconfinement at low SPS energies by NA49, and the establishment by NA61/SHINE of the diagram of high-energy nuclear collisions, featuring transitions between hadron-, string-, and QGP-dominated regimes. This predominantly scientific review is complemented by brief personal recollections related to the discussed topics.
\end{abstract}

\vspace{1em}

\section{NA35: first hints of QGP}
\label{sec:na35}

NA35 was the first experiment in the family of heavy-ion experiments conducted at the H2 beamline of the CERN North Area at the SPS. It collected data on collisions of light nuclear beams with various nuclear targets at 60$A$~GeV/$c$ and 200$A$~GeV/$c$ from 1986 to 1992. 
The main result of NA35 was the observation of enhanced production of strange hadrons in central S+S collisions at 200$A$~GeV/$c$. This marked the first indication of quark--gluon plasma (QGP) formation at the CERN SPS.  
During this period, I was a young postdoctoral researcher working on NA35 at universities in Heidelberg, Frankfurt am Main, and Warsaw.

\subsection{Background}
\label{subsec:na35_bkg}

The origins of the NA35 experiment at CERN can be traced back to a proposal~\cite{na35_roots} submitted by the GSI--LBL--Heidelberg--Marburg--Warsaw Collaboration to CERN's Proton Synchrotron and Synchro-Cyclotron Committee on January~26,~1982. The collaboration spokesperson was Reinhard Stock, a charismatic and enthusiastic leader of the streamer chamber heavy-ion experiment at the LBL Bevalac.

The proposal requested measurements of target fragmentation and hadron production in collisions of $^{16}$O with target nuclei ranging from $^{40}$Ca to $^{206}$Pb at the CERN Proton Synchrotron (PS). The requested beam energy range was 9$A$--13$A$~GeV/$c$, with an anticipated start of data taking in the spring of 1984, requiring 250~hours of dedicated PS running time. The ultimate physics goal was the observation of a hypothetical new state of matter, such as quark matter or quark--gluon plasma~\cite{Kapusta:1979fh,Shuryak:1980tp}, with explicit references to the work of Joseph Kapusta and Edward Shuryak included in the proposal.

Although CERN accepted the proposal, accommodating the experiment in the East Hall PS extraction area proved challenging. As a result, CERN management suggested~\cite{Klapisch:1984yfi} transporting the ion beam from the PS to the Super Proton Synchrotron (SPS) and distributing it to the North and West Halls of the SPS. The maximum SPS ion beam momentum of 200$A$~GeV/$c$ significantly expanded the programme's potential. This marked the beginning of CERN’s heavy-ion programme~\cite{Schukraft:2015dna} in general, and of NA35 in particular.

\subsection{Experiment}
\label{subsec:na35_exp}

\begin{figure}[htbp]
\begin{center}
\includegraphics[scale=0.40]{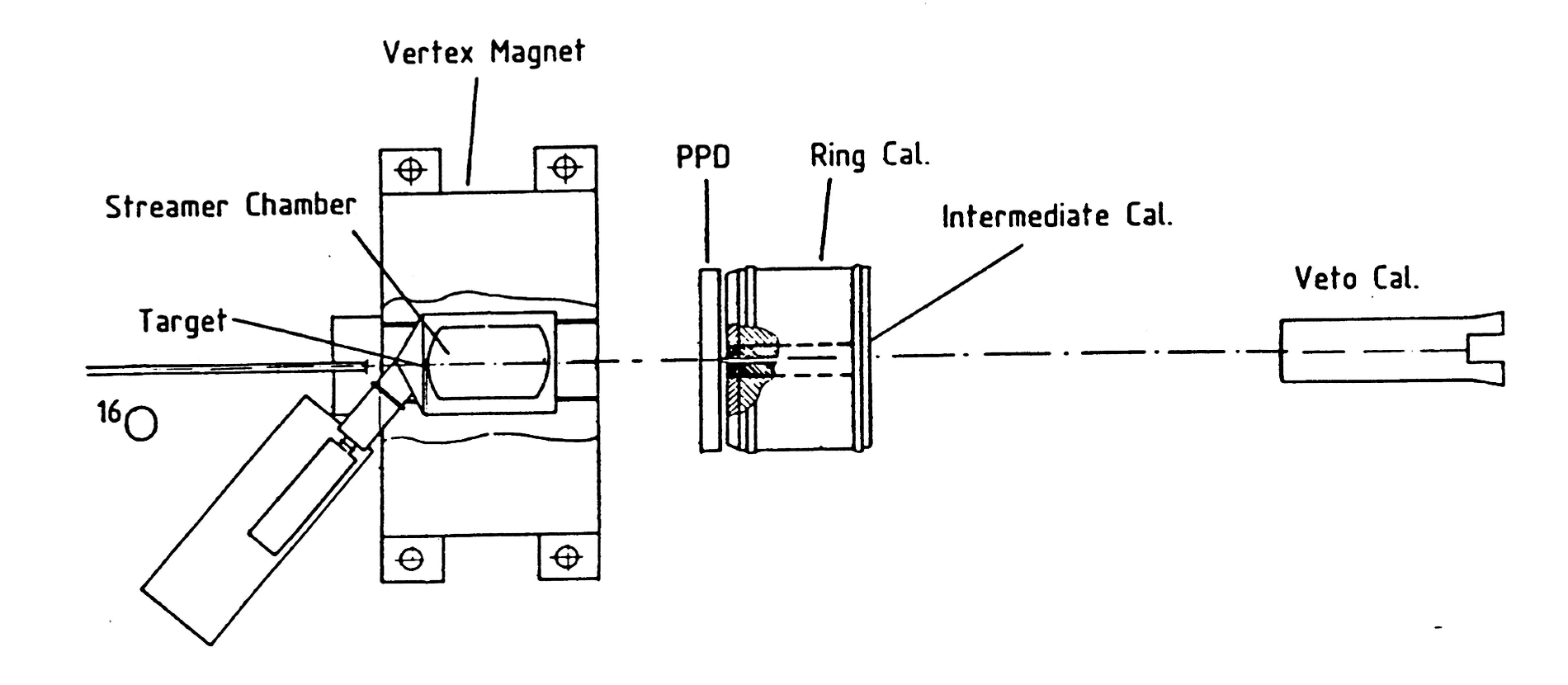}
\end{center}
\caption{
The top view of the NA35 experimental setup~\cite{Sandoval:170613}; see text for details. 
}
\label{fig:na35_setup}
\end{figure}

\begin{figure}[htbp]
\begin{center}
\includegraphics[scale=0.35]{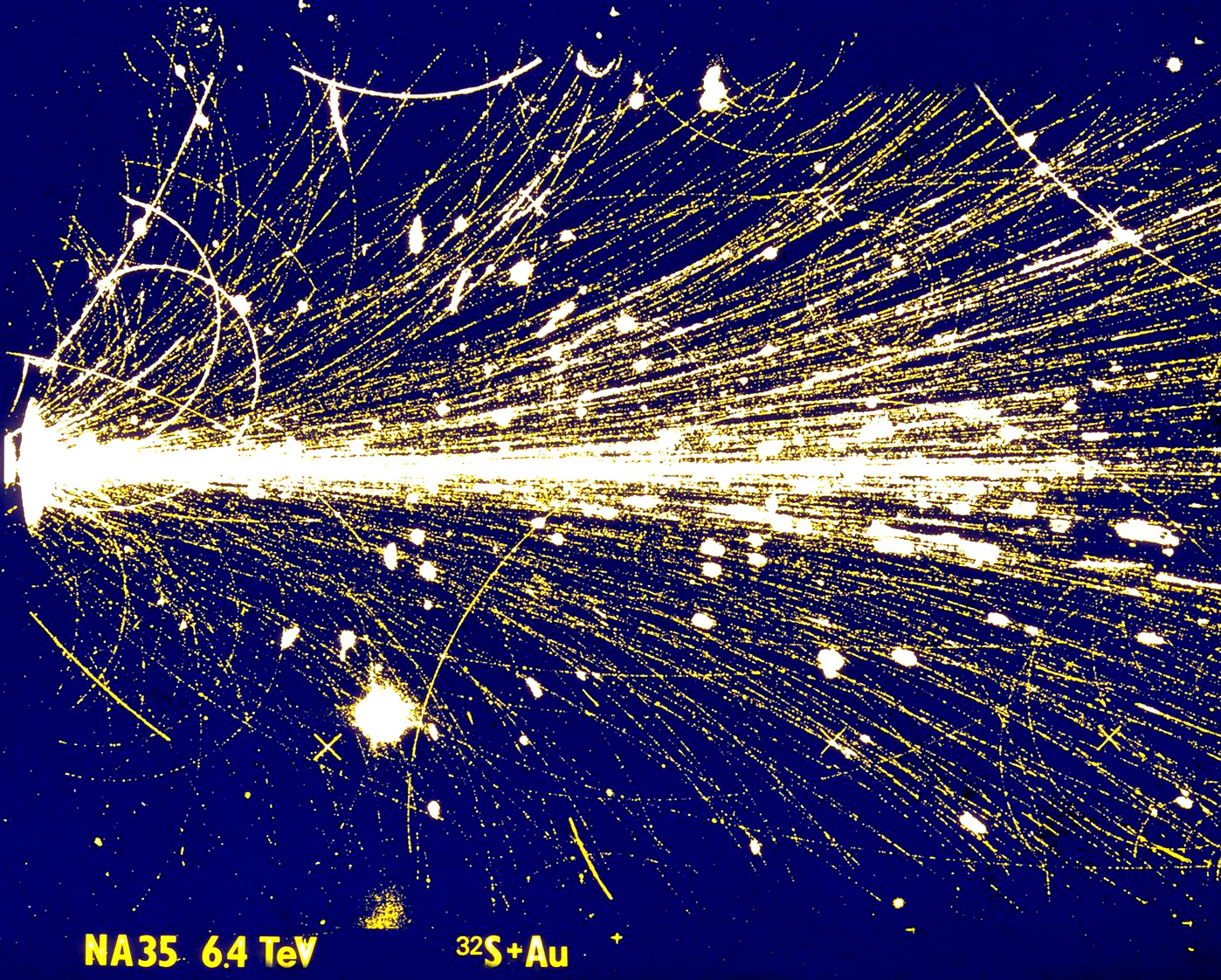}
\end{center}
\caption{
An iconic image of a central S+Au collision at 158$A$~GeV/$c$ recorded by the NA35 streamer chamber.
}
\label{fig:na35_event}
\end{figure}

The NA35 setup shown in Fig.~\ref{fig:na35_setup} featured a visual tracking detector, a streamer chamber with a sensitive volume of 200$\times$120$\times$70~cm$^3$, placed in a 1.5~T vertical magnetic field. Downstream, the detector was followed by four calorimeters. Major components of the experiment were inherited from the NA5 and NA24 experiments at the CERN SPS. For a detailed description, see Ref.~\cite{Sandoval:170613}. 
An iconic image of a central S+Au collision at 158$A$~GeV/$c$ is presented in Fig.~\ref{fig:na35_event}.

NA35 recorded data on collisions of $p$, $d$, $^{16}$O, and $^{32}$S beams at 60$A$~GeV/$c$ and 200$A$~GeV/$c$ with S, Ag, and Au targets from 1986 to 1992. 
In parallel, data with these beams were recoded by NA34~\cite{NA34:1984occ}, NA36~\cite{NA36:1992avc}, WA80~\cite{WA80:1995xza} and WA85/WA94~\cite{WA85:1991nsm,WA94:1995szb} experiments.

Initially, the collaboration was led by the experiment's initiator, Reinhard Stock. Later, when he shifted his focus to establishing the NA49 experiment, Peter Seyboth, a former leader of the NA5 particle physics experiment at the CERN SPS, took over as spokesperson. Among the many others who contributed to the success of NA35 were Robert Brockmann, Volker Eckardt, John Harris, Grazyna Odyniec, Rainer Renfordt, Andres Sandoval, Herbert Stroebele, and Sigfried Wenig.

\begin{figure}[!t]
\begin{center}
\includegraphics[scale=0.4]{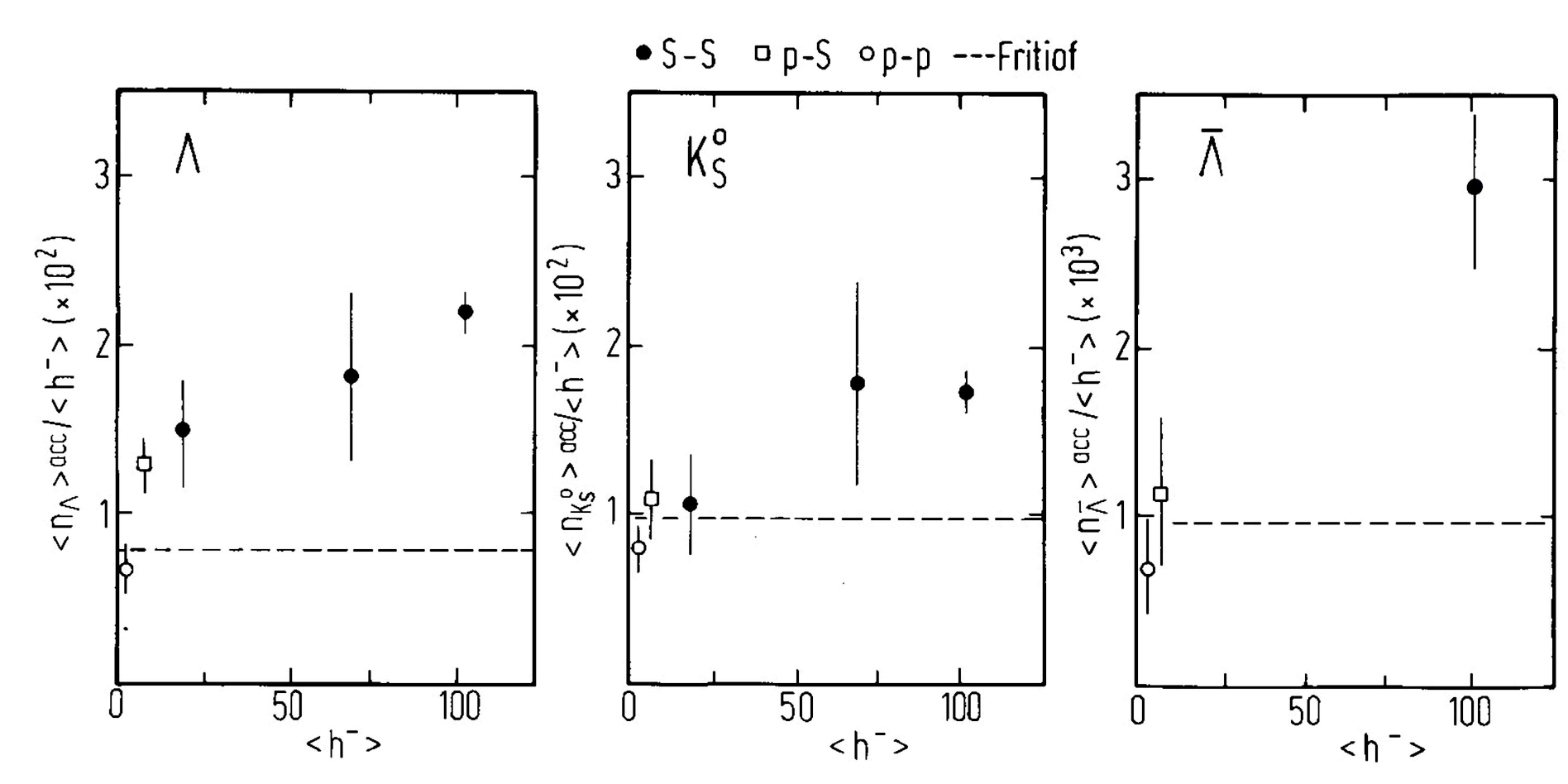}
\end{center}
\caption{
The ratio of the mean multiplicities of strange hadrons within the experimental acceptance to the total negative hadron multiplicity in S+S collisions at 200$A$~GeV/$c$ is shown for peripheral, intermediate, and central collisions. The two leftmost points represent the corresponding results for \textit{p+p} and $p$+S interactions.
The predictions of the simple superposition model, Fritiof, are independent of the $h^-$ mean multiplicity. The figure is taken from Ref.~\cite{NA35:1990teq}.
}
\label{fig:na35_strangeness}
\end{figure}

\subsection{Key results and conclusions}
\label{subsec:na35_res}

The most influential results of NA35 concerned neutral strange hadron production in central S+S collisions at 200$A$~GeV/$c$~\cite{NA35:1990teq}.  
Owing to the reflection symmetry of the initial state, it was possible to calculate the total mean multiplicity of abundant hadrons carrying strange and antistrange quarks ($s$ and $\bar{s}$), in particular $\Lambda$ hyperons and $K^0_S$ mesons. Their yields, relative to negatively charged hadrons ($h^-$, more than 90\% of which are $\pi^-$ mesons), were approximately twice as large as the corresponding yields in \textit{p+p} interactions. See Fig.~\ref{fig:na35_strangeness}, taken from Ref.~\cite{NA35:1990teq}, for illustration. 

This result indicated substantial production of $s\bar{s}$ pairs, consistent with the predictions of Johann Rafelski and Berndt M\"uller regarding quark--gluon plasma formation in such collisions~\cite{Rafelski:1982pu,Koch:1986ud}.

NA35 also achieved other important results, particularly on two-pion correlations~\cite{NA35:1988eto}. These results demonstrated that the pion--emission source is significantly larger than the size of the colliding nuclei, indicating expansion of the matter created in the early stages of the collisions.

\subsection{Memories}
\label{subsec:na35_mem}

I remember being in the beautiful garden of the GSI and TU Darmstadt guest house. While caring for my children, I scanned pages of printouts on fan-fold paper containing histograms and numbers from the final stage of the analysis of $\Lambda$ and $K^0_S$ production in S+S collisions. Eventually, I divided their yields by the $h^-$ yield and, to my shock, found ratios approximately twice as large as those observed in $p+p$ interactions. At that time, I did not believe in the possibility of quark--gluon plasma creation in nucleus--nucleus collisions and was convinced that I had made a mistake. However, subsequent checks by both me and the collaboration revealed no errors.

This discovery transformed me from a QGP sceptic into a believer. I presented the results~\cite{NA35:1989oqq} in 1988 at the Quark Matter conference in Lenox and shortly afterwards at the Hadronic Matter in Collision conference in Tucson, arguing that they were in quantitative agreement with QGP creation. The collaboration's paper was published in 1990~\cite{NA35:1990teq}.

\section{NA49-1: Event-by-event fluctuations and QGP}
\label{sec:na491}

NA49 is the second experiment in the family of heavy-ion experiments located at the H2 beamline of the CERN North Area at the SPS. The experiment encompassed two distinct physics programmes.
The first phase (NA49-1: 1994–1997), presented in this section, focused on the search for anomalies in event-by-event fluctuations in central Pb+Pb collisions at 158$A$~GeV/$c$, which could signal the creation of quark--gluon plasma (QGP). 
During this period, I was a scientific associate at the universities of Frankfurt am Main and Warsaw, working on NA49, STAR, and ALICE.

\subsection{Background}
\label{subsec:na491_bkg}

The NA35 Collaboration initiated the NA49 experiment with a proposal submitted~\cite{Panagiotou:295042} to the Super Proton Synchrotron and LEAR Committee in May 1991, which was approved in September 1991. The proposing institutions included groups from Athens, Berkeley, Darmstadt, Frankfurt am Main, Freiburg, Marburg, Munich, Seattle, Warsaw, and Zagreb. Reinhard Stock served as spokesperson. 

The primary aim of NA49 was to search for the deconfinement transition in central Pb+Pb collisions at 158$A$~GeV/$c$ by measuring event-by-event fluctuations of hadron production properties. This goal necessitated a significant upgrade of the NA35 detector to enable large-acceptance measurements of hadron production in high-multiplicity events (up to 1000 charged particles). Building on NA35's discovery of strangeness enhancement in S+S collisions~\cite{NA35:1990teq}, special emphasis was placed on the identification of charged kaons and pions.

\subsection{Experiment}
\label{subsec:na491_exp}

\begin{figure}[htbp]
\begin{center}
\includegraphics[scale=0.20]{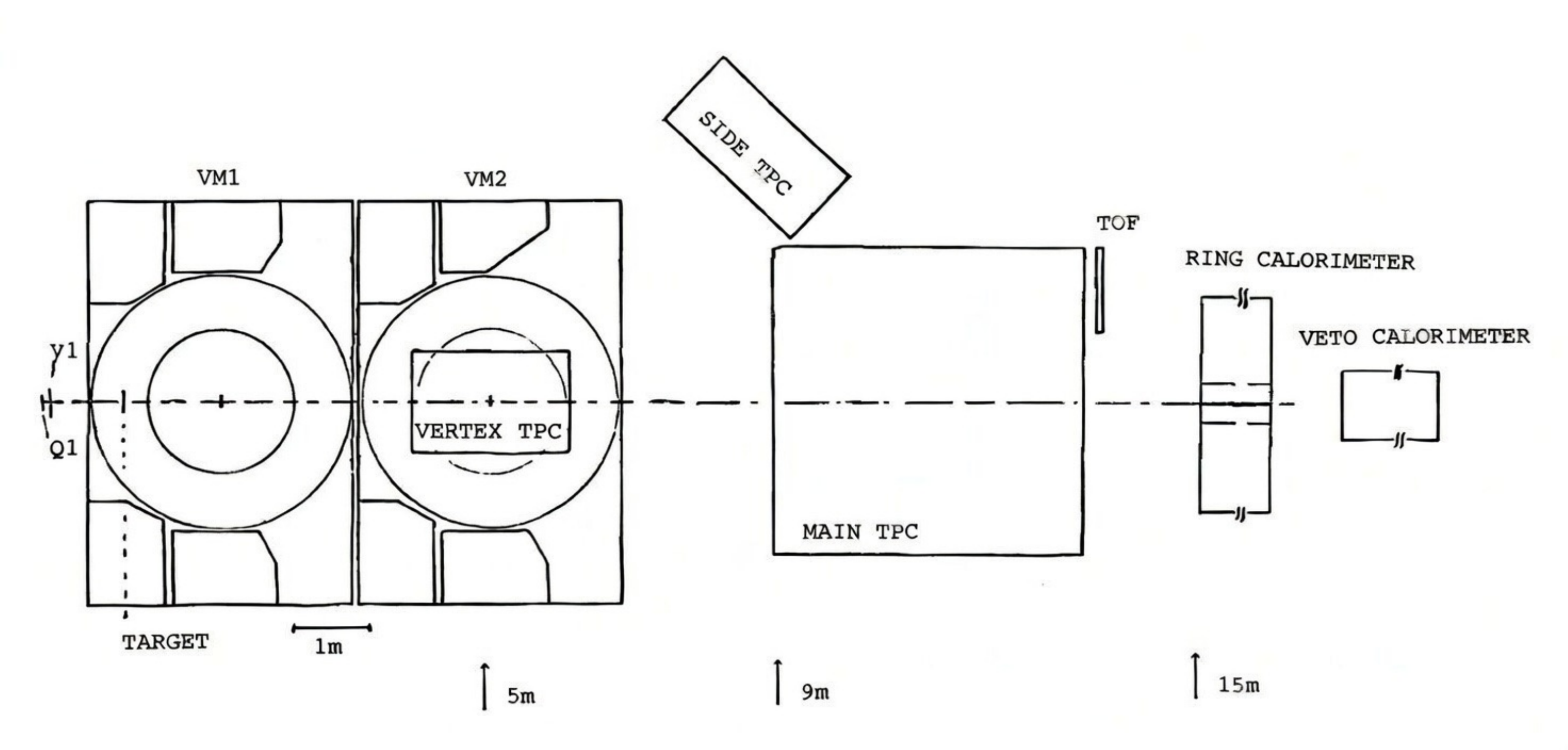}

\vspace*{0.5cm}
\includegraphics[scale=0.28]{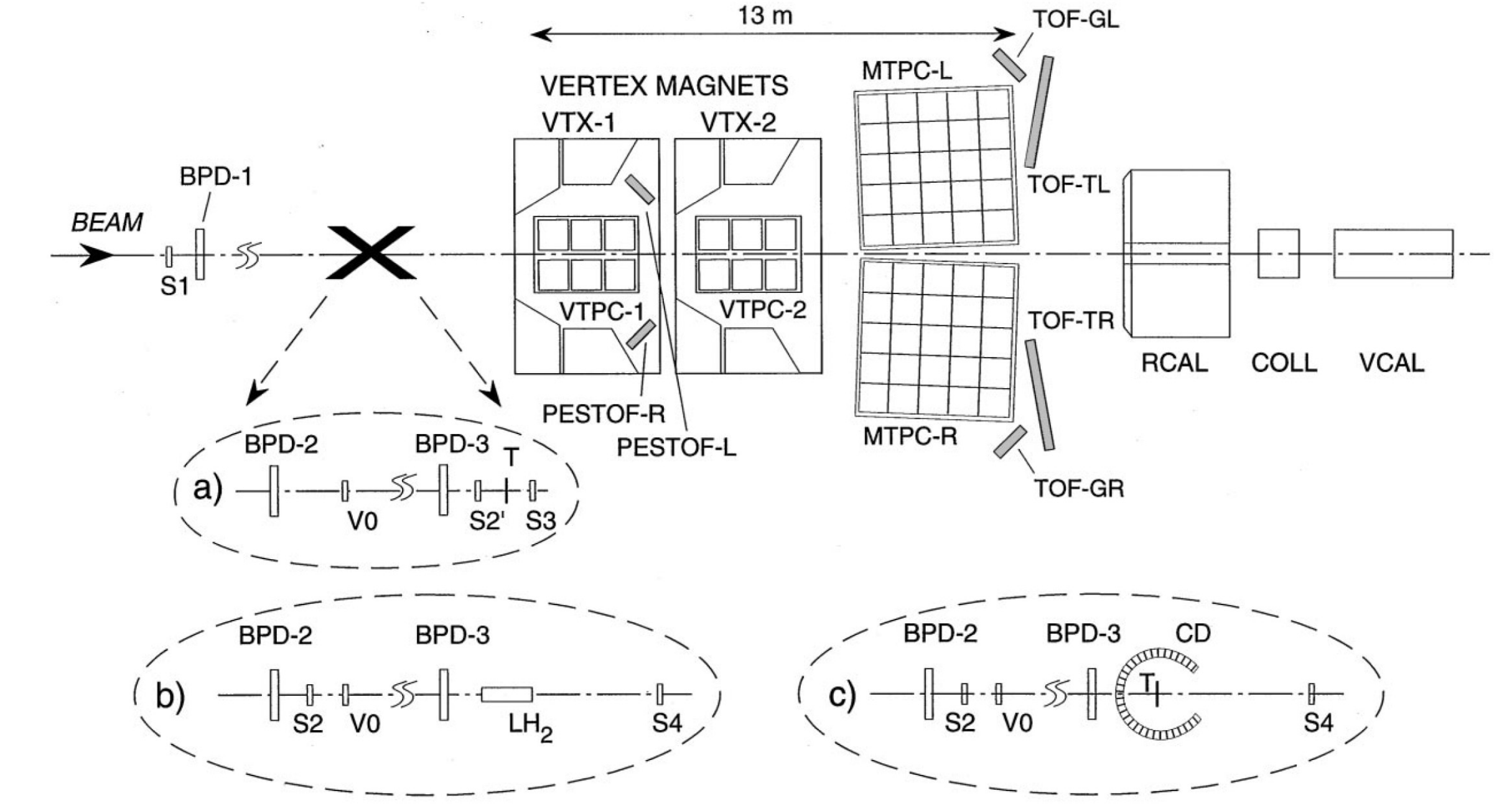}
\end{center}
\caption{
\textit{Top:} Top view of the NA49 experimental setup as originally proposed by NA35~\cite{Panagiotou:295042}.  
\textit{Bottom:} The setup constructed by the NA49 Collaboration~\cite{NA49:1999myq}.  
For further details, see the references above and the text.
}
\label{fig:na49_setup}
\end{figure}

The experimental setup proposed by the NA35 Collaboration, shown in Fig.~\ref{fig:na49_setup} (\textit{top}), underwent significant changes prior to the construction of the NA49 experiment. The final setup is depicted in Fig.~\ref{fig:na49_setup} (\textit{bottom}). 
The main components included four large-volume Time Projection Chambers (TPCs) for tracking and particle identification via energy loss in the TPC gas. Time-of-Flight (ToF) scintillator arrays extended charged-kaon identification down to mid-rapidity. Calorimeters for transverse-energy measurement and central-collision selection using forward energy completed the downstream acceptance. The figure illustrates the different configurations of beam and trigger detectors used for data taking in Pb+Pb~(a), 
\textit{p+p}~(b), and $p$+A~(c) collisions.

In 1995 and 1996, approximately 1.5 million central Pb+Pb collisions at 158$A$~GeV/$c$ were recorded, fulfilling the original NA49 data-taking plan. In 1997, Reinhard Stock transferred the role of spokesperson to Peter Seyboth.
In parallel, data with these beams were recoded by 
NA44~\cite{NA44:1996xlh}, NA45~\cite{CERESNA45:1997tgc}, NA50~\cite{NA50:2000brc}, NA52~\cite{NA52NEWMASS:1996uce} and WA97/NA57~\cite{WA97:1999uwz,NA57:2006aux} experiments.

\subsection{Key results and conclusions}
\label{subsec:na491_res}

The NA49 phase-one results were summarised by the collaboration in 1998 as follows~\cite{Bachler:356588}:  

\textit{
The analysis makes use of the unique capability of the NA49 detector to combine the study of particle yields with correlation analysis and physics on the event--by--event level. First results~\cite{NA49:1994hfj, NA49:1997qey, NA49:1997xvz, Roland:1997hs} shed new light on salient questions concerning the behaviour of matter at large energy density. In particular, they show that the transition from collisions of medium-size nuclei (as available in the past) to large-size systems does not lead to a substantial change in observables considered as possible signals of a phase transition, such as the pion-to-baryon and strangeness-to-pion ratios. A significant change in these observables is seen only when comparing nucleon--nucleon interactions with collisions of medium- and large-size nuclei at SPS energy. Our event--by--event analysis of Pb+Pb data does not reveal the appearance of dynamically different new event classes down to the permille level of admixture. 
}

The collaboration concluded~\cite{Bachler:356588}:
\textit{
These results, together with the energy dependence of pion and strangeness production in nucleus--nucleus (A+A) collisions, might indicate that the threshold energy for the creation of a deconfined partonic state (ideally the QGP) was already passed with SPS sulphur beams.} 
This conclusion marked the beginning of the second phase of NA49, which is discussed below.

\subsection{Memories}
\label{subsec:na491_mem}

Measuring and understanding event-by-event fluctuations of hadron production properties, such as the numbers of different hadron species produced in individual events, is a challenging task. I recall returning to Warsaw from a STAR Collaboration meeting in Berkeley. During the long journey and subsequent jet lag, I contemplated the possibility of separating dynamical and geometrical fluctuations in Pb+Pb collisions. Since both types of fluctuations were unknown, the task initially seemed hopeless. Surprisingly, by using a minimal model of the problem, I guessed an analytical solution that later turned out to be correct. 

Early the next morning, I rushed to the institute to verify the solution using a simple Monte Carlo simulation (at that time, I had neither a PC nor a laptop at home). The solution passed the numerical check. Shortly thereafter, Staszek Mrowczynski derived an analytical proof of the conjectured solution, and we published the results~\cite{Gazdzicki:1992ri}. This marked the beginning of the strongly intensive quantities~\cite{Gorenstein:2011vq} approach to fluctuation studies.

I applied the same problem-solving strategy—guessing the analytical solution and asking Staszek to prove it—when proposing a method for fluctuation studies with incomplete hadron identification~\cite{Gazdzicki:2011xz}. The so-called identity method was later significantly developed~\cite{Gorenstein:2011hr, Rustamov:2012bx} and has since been widely used.

\newpage
\section{NA49-2: Discovery of the onset of deconfinement}
\label{sec:na492}

The second phase of NA49 (NA49-2: 1998--2002), presented here, was devoted to the search for the onset of QGP creation in Pb+Pb collisions by reducing the collision energy. The primary result was the observation of anomalies---referred to as the horn, kink, and step---in the collision-energy dependence of various hadron production properties. These findings suggest that QGP creation begins at low SPS beam momenta, around 30$A$~GeV/$c$.\\
During this period, I was a scientific associate at the University of Frankfurt am Main and at CERN, working mostly on NA49.

\subsection{Background}
\label{subsec:na492_bkg}

In the mid-1990s, numerous results were obtained from collisions of light nuclei at the BNL AGS (beams of Si at 14.6$A$~GeV) and the CERN SPS (beams of O and S at 200$A$~GeV). Experiments with heavy nuclei (AGS: Au+Au at 11.6$A$~GeV; SPS: Pb+Pb at 158$A$~GeV) were only just beginning. This period marked the first opportunity to examine the energy dependence of hadron production in nucleus--nucleus collisions at high energies. 

Two compilations---one on pion production~\cite{Gazdzicki:1995zs} and another on strangeness production~\cite{Gazdzicki:1996pk}---led to a clear conclusion: the energy dependence of mean hadron multiplicities measured in A+A collisions differs significantly from that observed in \textit{p+p} interactions. Moreover, the A+A data suggested a notable change in the energy dependence of pion and strangeness yields, located between the top AGS and SPS energies. 

Using a statistical approach to strong interactions~\cite{Fermi:1950frz, Landau:1953wku}, I speculated~\cite{Gazdzicki:1995ze} that this change is related to the onset of deconfinement in the early stage of A+A collisions. Following this conjecture, a quantitative model, the Statistical Model of the Early Stage (SMES)~\cite{Gazdzicki:1998vd}, was developed in collaboration with Mark Gorenstein. The SMES assumes the creation of early-stage matter according to the principle of maximum entropy. Depending on the collision energy, the matter exists in the confined phase ($E \lesssim 30$~$A$~GeV), the mixed phase ($30 \lesssim E \lesssim 60$~$A$~GeV), or the deconfined phase ($E \gtrsim 60$~$A$~GeV). The phase transition is assumed to be of first order.

In 1997, based on these ideas, the NA49 Collaboration proposed studying hadron production in Pb+Pb collisions at 40$A$~GeV~\cite{Bachler:356588}, marking the beginning of the second phase of the NA49 experiment.

\subsection{Experiment}
\label{subsec:na492_exp}

The NA49-2 programme was conducted using essentially the same experimental setup as NA49-1; see Sec.~\ref{subsec:na491_exp}. The only modification was to scale the magnetic field in proportion to the beam momentum to ensure kaon identification at mid-rapidity via time-of-flight and specific energy-loss measurements in the TPCs; see Fig.~\ref{fig:na49_setup}.

The first 40$A$~GeV Pb beam was delivered to NA49-2 in 1998 as a test. A five-week run at 40$A$~GeV followed in 1999. The success of this initial run at low SPS energy, combined with the promising preliminary results presented by NA49, justified continuation of the programme. In 2000, a beam at 80$A$~GeV was delivered to NA49 for a five-day run. 
The programme was completed in 2002 with two additional runs: one at 30$A$~GeV (seven days) and another at 20$A$~GeV (seven days).
In parallel, data with these beams were recoded by the NA45~\cite{CERESNA45:2002gnc} 
and NA60~\cite{NA60:2006ymb} experiments.

\subsection{Key results and conclusions}
\label{subsec:na492_res}

The basic NA49-2 results concerning the onset of deconfinement were published in 2002~\cite{NA49:2002pzu} and 2008~\cite{NA49:2007stj}. These results, focusing on pion and kaon production in central Pb+Pb collisions at 40$A$~GeV, 80$A$~GeV, 158$A$~GeV, and later at 20$A$~GeV and 30$A$~GeV, have together accumulated about 1300 citations.

The collaboration concluded~\cite{NA49:2007stj}:
\textit{
A rapid change of the energy dependence is observed around 30$A$~GeV for the yields of pions and kaons as well as for the shape of the transverse mass spectra. The change is compatible with the prediction that the threshold for production of a state of deconfined matter at the early stage of the collisions is located at low SPS
energies.
}

Following Ref.~\cite{NA49:2007stj}, I briefly report the two most striking results and their relation to the onset of deconfinement, noting that many other results from the beam-energy scan of identified hadron yields, correlations, and fluctuations have also been published.

The $\langle K^+ \rangle / \langle \pi^+ \rangle$ and $E_S = (\langle \Lambda \rangle + 4 \langle K^0_S \rangle)/ \langle \pi \rangle$ ratios are approximately proportional to the total strangeness-to-entropy ratio, which in the SMES model~\cite{Gazdzicki:1998vd} is assumed to remain constant from the early stage to freeze-out. At low collision energies, the strangeness-to-entropy ratio increases steeply with energy owing to the low temperature at the early stage ($T \lesssim T_C$) and the high mass of the strangeness carriers in the confined state (e.g.\ the kaon mass of 500~MeV). 

\begin{figure}[htbp]
\begin{center}
\includegraphics[scale=0.48]{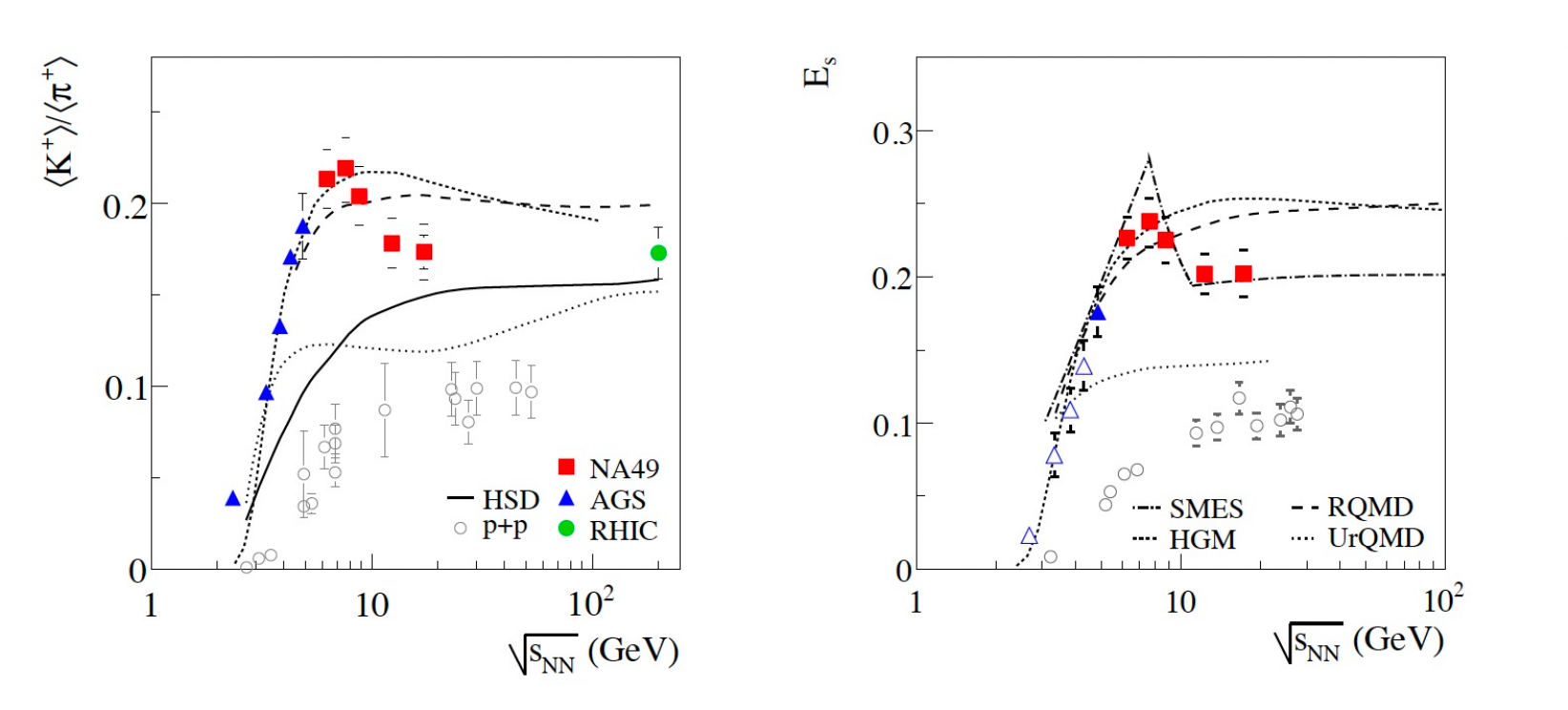}
\end{center}

\caption{
\textit{Left:} Energy dependence of the $\langle K^+ \rangle / \langle \pi^+ \rangle$ ratio measured in central Pb+Pb and Au+Au collisions (full symbols), compared to the corresponding results from \textit{p+p}($\bar{p}$) reactions (open circles).  
\textit{Right:} Energy dependence of relative strangeness production as measured by the $E_S$ ratio in central Pb+Pb and Au+Au collisions (full symbols), compared to results from \textit{p+p}($\bar{p}$) reactions (open circles). The curves in the figures represent predictions from various models. 
The plots and caption are taken from Ref.~\cite{NA49:2007stj}, where details and references are provided.
}
\label{fig:na49_horn}      
\end{figure}

When the transition to a QGP is crossed ($T \gtrsim T_C$), the mass of the strangeness carriers drops significantly to the strange-quark mass of about 100~MeV. As a result, when $m < T$, the strangeness yield becomes approximately proportional to the entropy, and the strangeness-to-entropy (or pion) ratio becomes independent of energy. This transition leads to a decrease in the energy dependence of the ratio from the larger value for confined matter at $T_C$ to the QGP value. Consequently, the measured non-monotonic energy dependence of the strangeness-to-entropy ratio is followed by saturation at the QGP value. The corresponding anomalous energy dependence of the $\langle K^+ \rangle / \langle \pi^+ \rangle$ and $E_S$ ratios, referred to as the ``horn'', is shown in Fig.~\ref{fig:na49_horn} and, within the SMES framework, is a direct consequence of the onset of deconfinement occurring at around 30$A$~GeV.

In the mixed-phase region, the early-stage pressure and temperature are independent of the energy density~\cite{VanHove:1982vk}. Consequently, within the SMES model, this results in a weakening of the increase with energy of the inverse slope parameter of transverse mass spectra (T: $1/m_T dn/dm_T \sim e^{-m_T/\textrm{T}}$) (the ``step'') or, equivalently, the mean transverse mass $\langle m_T \rangle$ in the SPS energy range~\cite{Gorenstein:2003cu}. 

\begin{figure}[htbp]
\begin{center}
\includegraphics[scale=0.45]{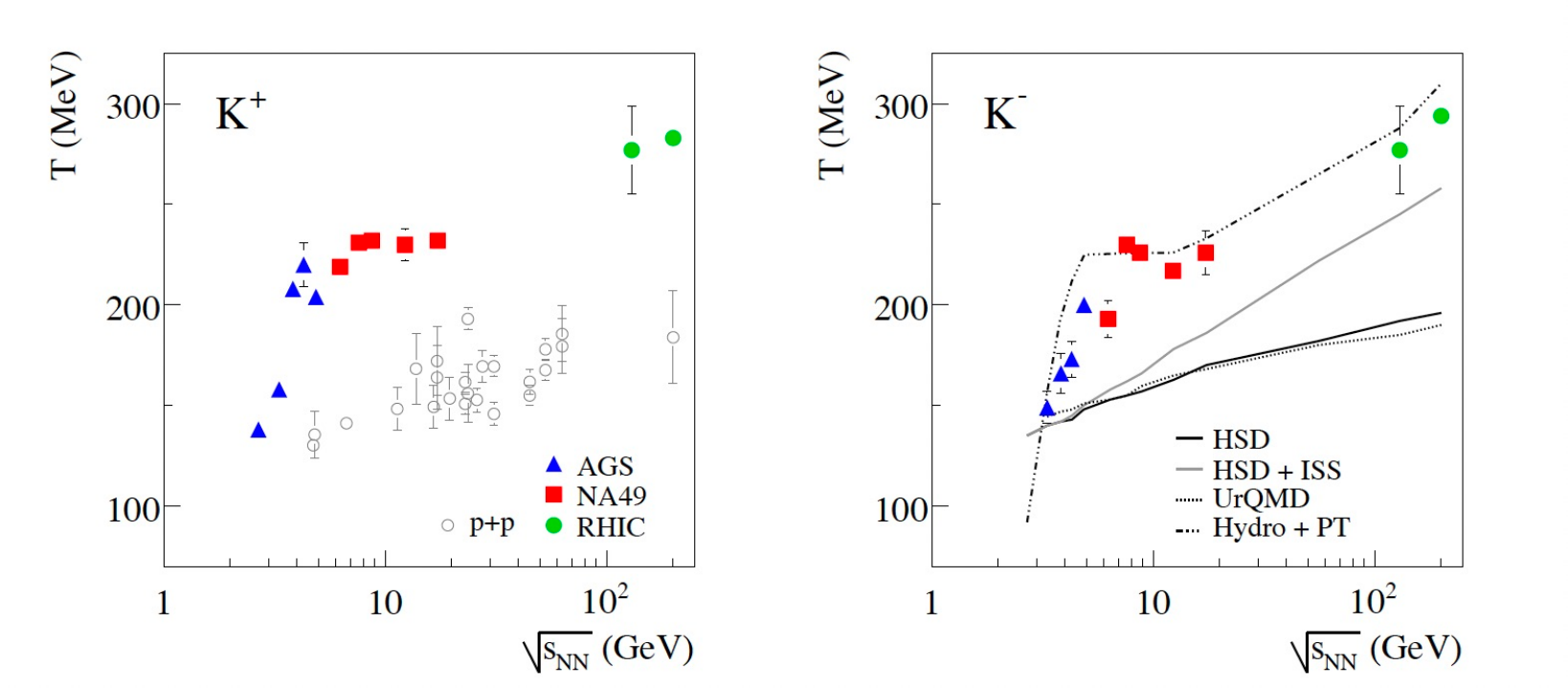}
\end{center}

\caption{
Energy dependence of the inverse slope parameter $T$ of the transverse mass spectra for $K^+$ (\textit{left}) and $K^-$ mesons (\textit{right}) measured at mid-rapidity in central Pb+Pb and Au+Au collisions. The $K^+$ slope parameters are compared 
to those from \textit{p+p}($\bar{p}$) reactions in the left-hand plot (open circles). The curves in the right-hand plot represent predictions from various models. 
The plots and caption are taken from Ref.~\cite{NA49:2007stj}, where details and references are provided.
}
\label{fig:na49_step}      
\end{figure}

This qualitative prediction is confirmed by the results shown in Fig.~\ref{fig:na49_step}, where the collision-energy dependence of the T parameter is shown for heavy-ion and $p+p$ collisions
(\textit{left}) and compared with model predictions (\textit{right}). The hydrodynamic calculations~\cite{Gazdzicki:2003dx} that incorporate the transition between hadronic and deconfined phases (labelled as "Hydro+PT" in Fig.~\ref{fig:na49_step} (\textit{right)}) describe the data, whereas the other models, having no phase transition, fail to reproduce the experimental results.

The observation of the predicted signals of deconfinement (the horn, step, and kink; see above and Sec.~\ref{subsec:na492_mem}) and the markedly different collision-energy dependence of hadron production properties in central Pb+Pb collisions compared to \textit{p+p} interactions underscored the need to continue the heavy-ion programme at the CERN SPS and other facilities worldwide. This motivated several initiatives, including NA61/SHINE at the CERN SPS (see the next section), the Beam Energy Scan programme~\cite{STAR:2010vob} at the Relativistic Heavy Ion Collider (RHIC) at Brookhaven National Laboratory, the CBM experiment~\cite{CBM:2016kpk} at the Facility for Antiproton and Ion Research (FAIR) in Darmstadt, and the MPD experiment~\cite{MPD:2022qhn} at NICA in Dubna.  

A series of workshops titled \textit{Critical Point and Onset of Deconfinement} was established in 2004 by me, Peter Seyboth, and Edward Shuryak at the European Centre for Theoretical Studies in Nuclear Physics and Related Areas (ECT*), Trento~\cite{CPOD2024}. During the first workshop in Trento, the search for the critical point of strongly interacting matter was identified as a key objective of these new programmes.

\subsection{Memories}
\label{subsec:na492_mem}

I remember walking across the endless car parks of Frankfurter Messe, where the institute was located at the time, contemplating the following puzzle: results on strangeness production indicate that QGP is created in central S+S collisions at the top SPS collision energy. Models suggest that the transition to QGP at $T = T_C$ should result in an entropy increase by a factor of about three. Since the pion number is approximately proportional to the entropy, one would naively expect the number of pions in nucleus--nucleus collisions to increase by $\approx 3$ when QGP starts to form. However, experimental data do not show this increase~\cite{Gazdzicki:1995zs}. What could be wrong?

Before entering the office, I found a rather obvious answer: the comparison of entropy between QGP and hadron gas relevant for nucleus--nucleus collisions should be done at fixed collision energy (or, equivalently, fixed energy density), not at fixed temperature. When this approach is applied, the pion number is expected to increase by a factor of $\approx 3^{1/4} \approx 1.3$, rather than 3~\cite{Gazdzicki:1995ze}. With this correction, the data are consistent with the onset of QGP creation at low SPS energies.

Eager to share the solution with my colleagues, I changed the subject of my talk~\cite{Gazdzicki:1994ud} at the last moment during the Divonne-les-Bains workshop dedicated to Rolf Hagedorn on the occasion of his 75th birthday in June 1994. The following handwritten letter from Rolf Hagedorn reads~\cite{Gazdzicki:2015oya}:

\textit{
Dear Dr. Ga\'zdzicki, \\
It was a pleasure to listen to your presentation of entropy production in nuclear collisions. It would be interesting to see your interpretation of overproduction 
in S+S collisions at 200~GeV/c/A confirmed by future data and by independent theoretical estimates of the change of entropy at the phase transition.
A lot of work remains to be done before we know there is a (one, may be two?) phase
transition and what are the details. Good luck for the future! Yours Rolf Hagedorn
}

Hagedorn was right: the talk in Divonne-les-Bains marked the first step in a long journey to study the onset of deconfinement in heavy-ion collisions. After 30 years of impressive progress, we have learned a great deal, yet many ``details'' remain to be established.

\section{NA61/SHINE: Diagram of high-energy nuclear collisions}
\label{sec:na61}

NA61/SHINE is the third and most recent experiment in the family of heavy-ion experiments located at the H2 beamline of the CERN North Area at the SPS. The experiment was approved by the CERN Research Board in 2007. Measurements for strong interaction physics are conducted in parallel with reference measurements for neutrino and cosmic-ray physics. The first phase of the strong interaction programme includes traditional heavy-ion physics. The part motivated by the quark-gluon plasma-related results of NA35 and NA49 is presented here. \\
The experiment conducted the world's first systematic scan in the two-dimensional space of laboratory-controlled parameters: the mass of colliding nuclei and collision energy. Results from \textit{p+p} and Be+Be collisions indicate a transition from hadron production dominated by resonances to that dominated by strings at $\sqrt{s_{NN}} \approx 10$~GeV. Moreover, results at the top SPS energy reveal a transition from string-dominated to quark-gluon plasma-dominated production between Be+Be and Ar+Sc collisions. These findings, combined with results from NA49, STAR, and ALICE, enable the establishment of the diagram of high-energy nuclear collisions.\\
During this period, I was a professor at Jan Kochanowski University in Kielce and a scientific associate at Goethe University Frankfurt. I was the founder and served as the spokesperson of NA61/SHINE for about 20 years.

\subsection{Background}
\label{subsec:na61_bkg}

\begin{figure}[htbp]
\begin{center}
\includegraphics[scale=0.15]{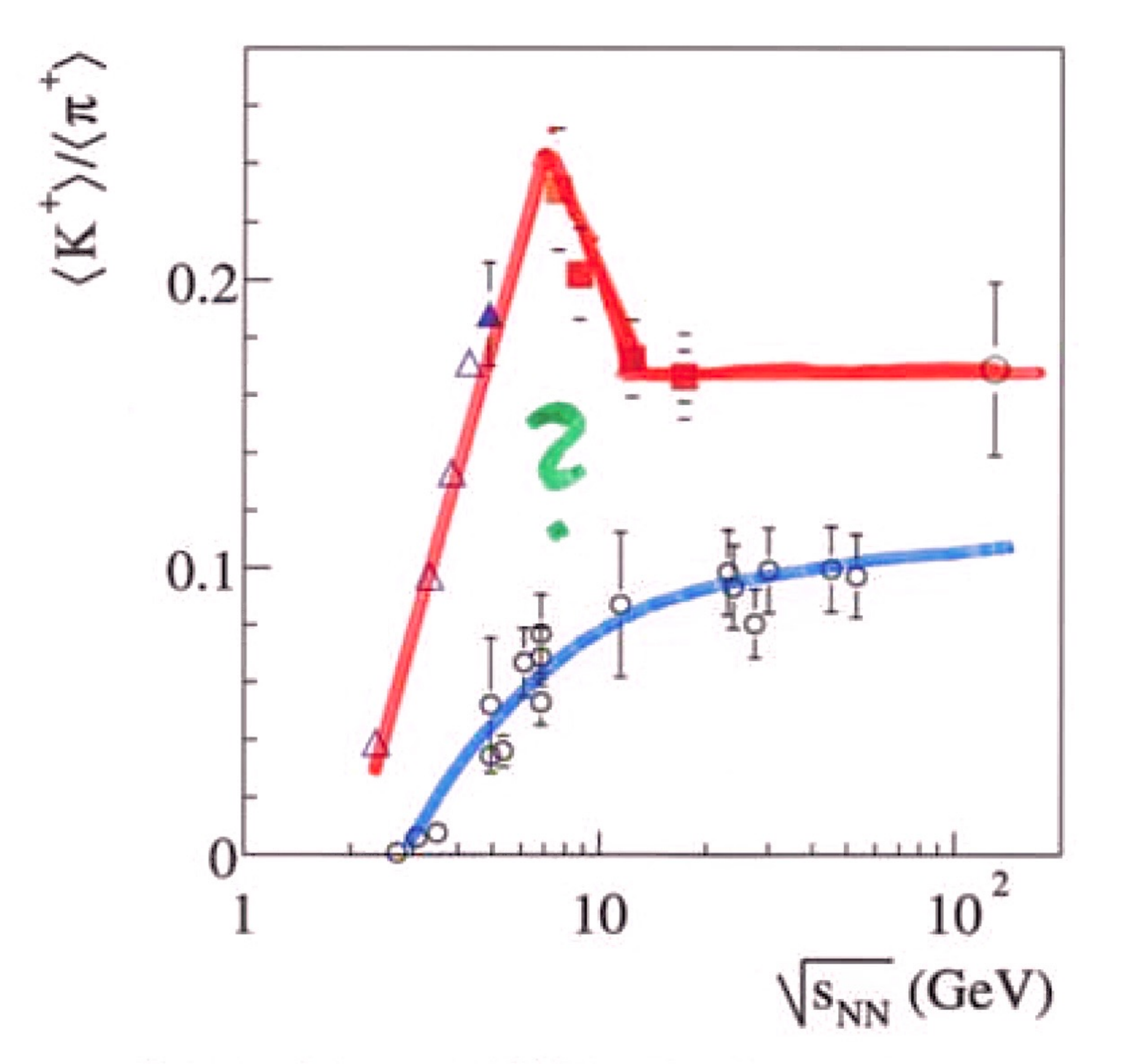}
\raisebox{1cm}{
\includegraphics[scale=0.15]{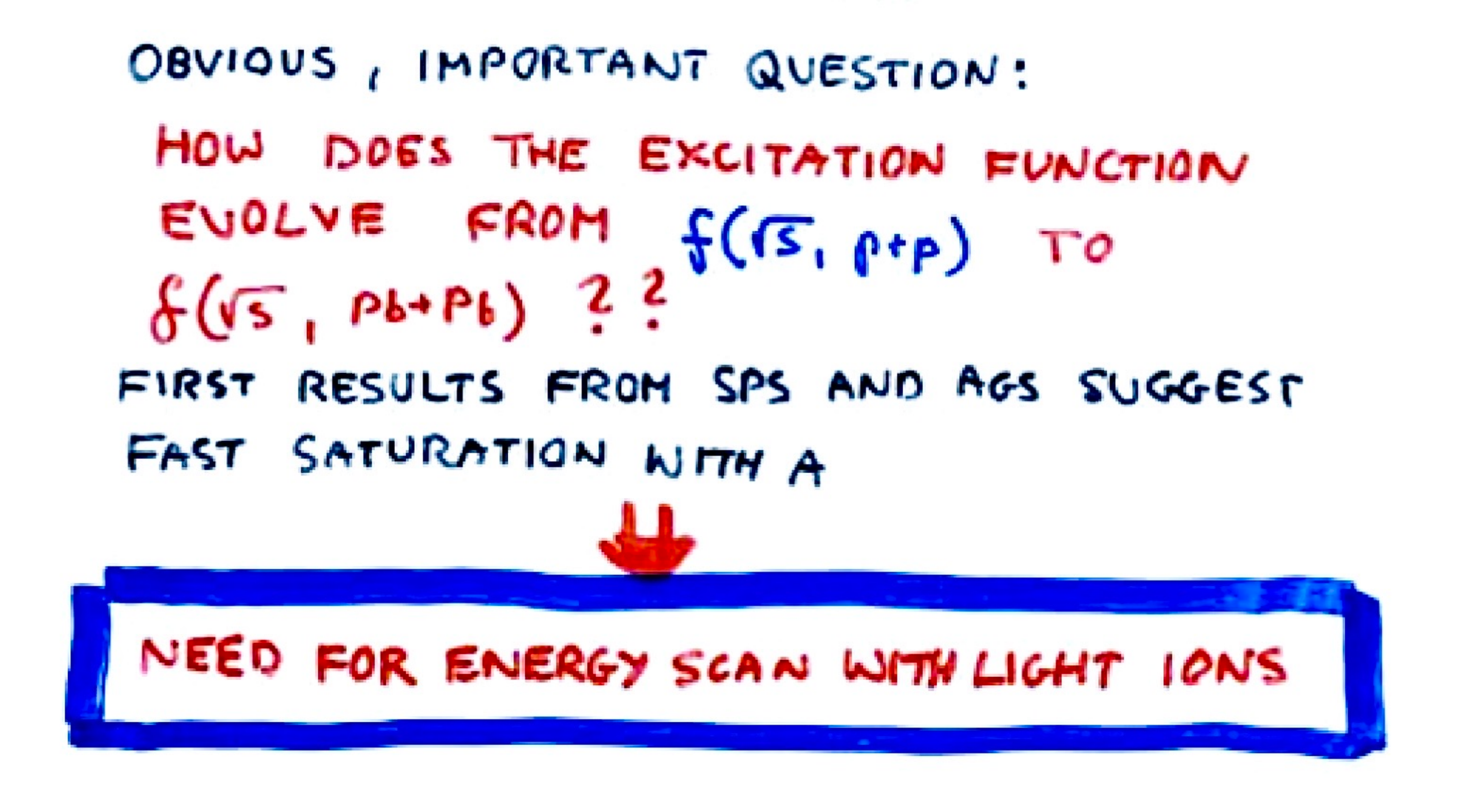}
}
\end{center}

\caption{
Scanned slide from my presentation at the NA49 Collaboration meeting at CERN in March 2003, which initiated NA61/SHINE~\cite{Gazdzicki:1998vd}.
}

\label{fig:na61_bkg}      
\end{figure}

The need for a two-dimensional scan of collision energy and colliding-nuclei masses at the CERN SPS was presented by me at the NA49 Collaboration meeting in March 2003; see Fig.~\ref{fig:na61_bkg} showing
the scan of the key slide. The idea was based on results from the NA49 energy scan with Pb+Pb collisions, which revealed a markedly different collision energy dependence observed for \textit{p+p} interactions and central Pb+Pb collisions; see Fig.~\ref{fig:na61_bkg}. Importantly, the dependence for Pb+Pb collisions aligns with the predictions for the onset of deconfinement at low SPS energies~\cite{Gazdzicki:1998vd}. 

The proposal for these new measurements was supported by the NA49 spokesperson, Peter Seyboth, and a group of NA49 senior members, including Zoltan Fodor (Budapest), Wojtek Dominik (Warsaw), Rainer Renford (Frankfurt am Main), Herbert Stroebele (Frankfurt am Main), and George Vesztergombi (Budapest).

At the Research Board meeting in June 2003, Agnieszka Zalewska, the Board Chair, remarked that a detailed study of the low-energy region would be challenging at RHIC and encouraged the preparation of ideas for further studies of heavy ions at the CERN SPS beyond 2005. 

The Expression of Interest, titled "A New Experimental Programme with Nuclei and Proton Beams at the CERN SPS," was submitted to the SPSC in November 2003~\cite{Gazdzicki:685283}, followed by the Letter of Intent in January 2006~\cite{Gazdzicki:919966}. The physics goals included a scan of the mass of colliding nuclei and collision energy, a search for the critical point of strongly interacting matter, and reference measurements for neutrino and cosmic-ray physics. Finally, the proposal, "Study Hadron Production in Hadron-Proton Interactions and Nucleus-Nucleus Collisions at the CERN SPS," was submitted to the SPSC in November 2006~\cite{Gazdzicki:995681}.

The scan programme received significant support from the theory community. Notably, Frank Wilczek, Edward Shuryak, Krishna Rajagopal, and Misha Stephanov sent letters and emails of endorsement. For further details, see the NA61/SHINE website~\cite{NA61_history}.

The new experiment, NA61/SHINE, was approved by the CERN Research Board in February 2007.

\subsection{Experiment}
\label{subsec:na61_exp}

\begin{figure}[htbp]
\begin{center}
\includegraphics[scale=0.35]{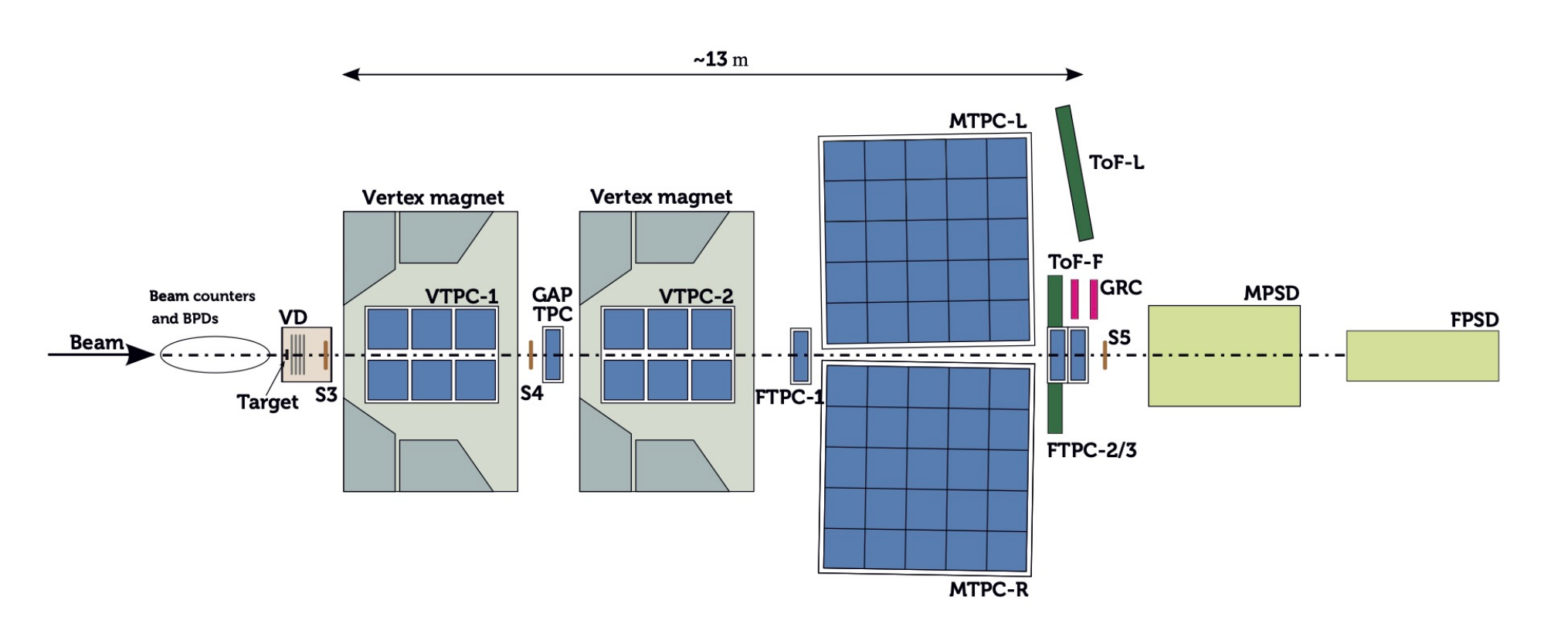}
\includegraphics[scale=0.20]{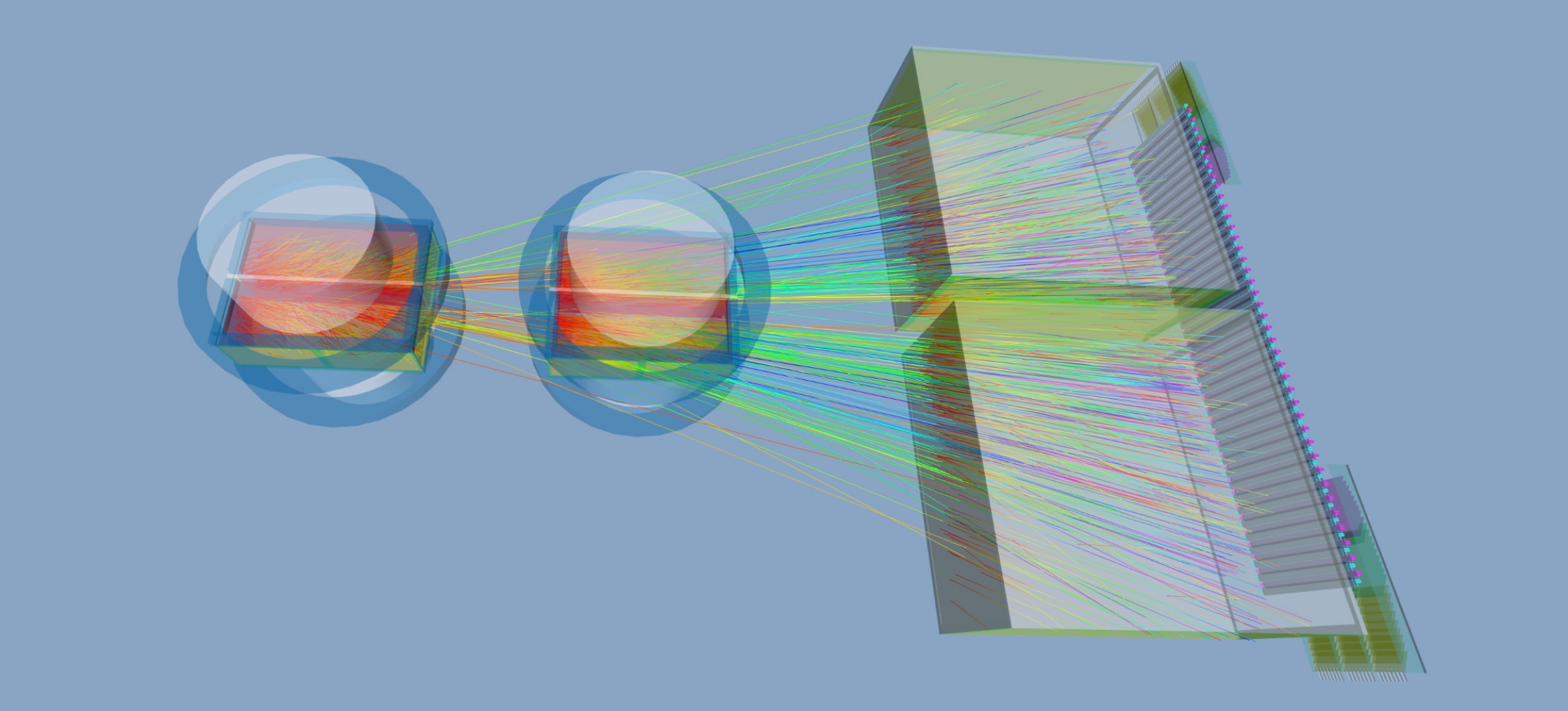}
\end{center}

\caption{
\textit{Top:} The post-LS2 NA61/SHINE detector~\cite{vonDoetinchem:2914265}.  
\textit{Bottom:} Particle tracks produced in an Xe+La collision and recorded by the NA61/SHINE Time Projection Chambers.
}

\label{fig:na61_postLS2}      
\end{figure}

The NA49 collaboration transferred ownership of the NA49 detector to NA61/SHINE. Following a successful test in 2006, pilot physics data-taking for the long-baseline neutrino experiment, T2K at J-PARC, began in 2007~\cite{NA61SHINE:2011dsu}. Since then, the detector has been continuously upgraded, with major upgrades implemented during accelerator shutdowns.

Data collection for the 2D scan programme was conducted from 2009 to 2017, utilising an evolving detector setup closely resembling the one presented in Ref.~\cite{NA61:2014lfx}. The Run~3 setup~\cite{vonDoetinchem:2914265} is shown in Fig.~\ref{fig:na61_postLS2}~(\textit{top}). The main components of the tracking system in the large-acceptance hadron spectrometer include a silicon pixel detector (VD) located downstream of the target, followed by four large-volume Time Projection Chambers (TPCs). Two Vertex TPCs (VTPC-1/2) are located within superconducting magnets, with a maximum combined bending power of 9 Tm. The magnetic field was scaled in proportion to the beam momentum to maintain similar rapidity-transverse momentum acceptance at all beam momenta. The main TPCs (MTPC-L/R) and Time-of-Flight (ToF) detectors are located downstream of the magnets. The setup is completed by two forward calorimeters, the Projectile Spectator Detectors (MPSD and FPSD). An event--browser display showing particle tracks produced in an Xe+La collision and recorded by the TPCs is presented in Fig.~\ref{fig:na61_postLS2}~(\textit{bottom}).

Data with were recorded for \textit{p+p} (secondary proton beams), Be+Be (secondary $^7$Be beams), Ar+Sc, and Xe+La collisions at 13$A$~GeV/$c$, 19(20)$A$~GeV/$c$, 30$A$~GeV/$c$, 40$A$~GeV/$c$, 75(80)$A$~GeV/$c$, and 150(158)$A$~GeV/$c$, where the number represents the laboratory beam momentum per nucleon. Additionally, data for Pb+Pb collisions were recorded at 13$A$~GeV/$c$, 30$A$~GeV/$c$, and 150$A$~GeV/$c$. The recorded datasets are summarised in Fig.~\ref{fig:na61_boxplot}~(\textit{left}).
In parallel, data  with ion beams were recoded by the CERN LHC experiments, 
ALICE~\cite{ALICE:2008ngc}, ATLAS~\cite{ATLAS:2014ipv} and CMS~\cite{CMS:2015yux}.

\subsection{Key results and conclusions}
\label{subsec:na61_res}\begin{figure}[htbp]

\begin{center}
\includegraphics[scale=0.18]{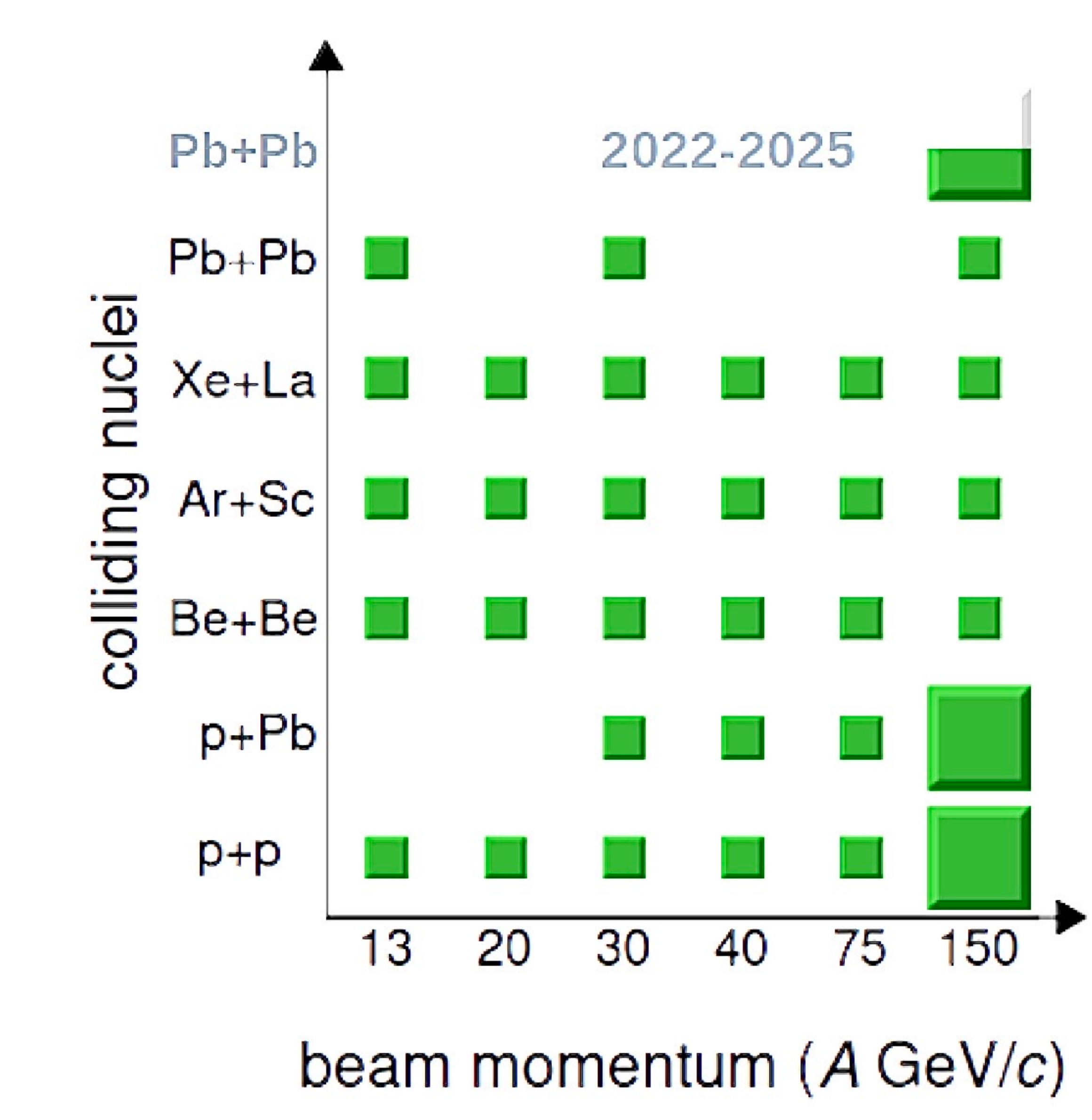}
\includegraphics[scale=0.50]{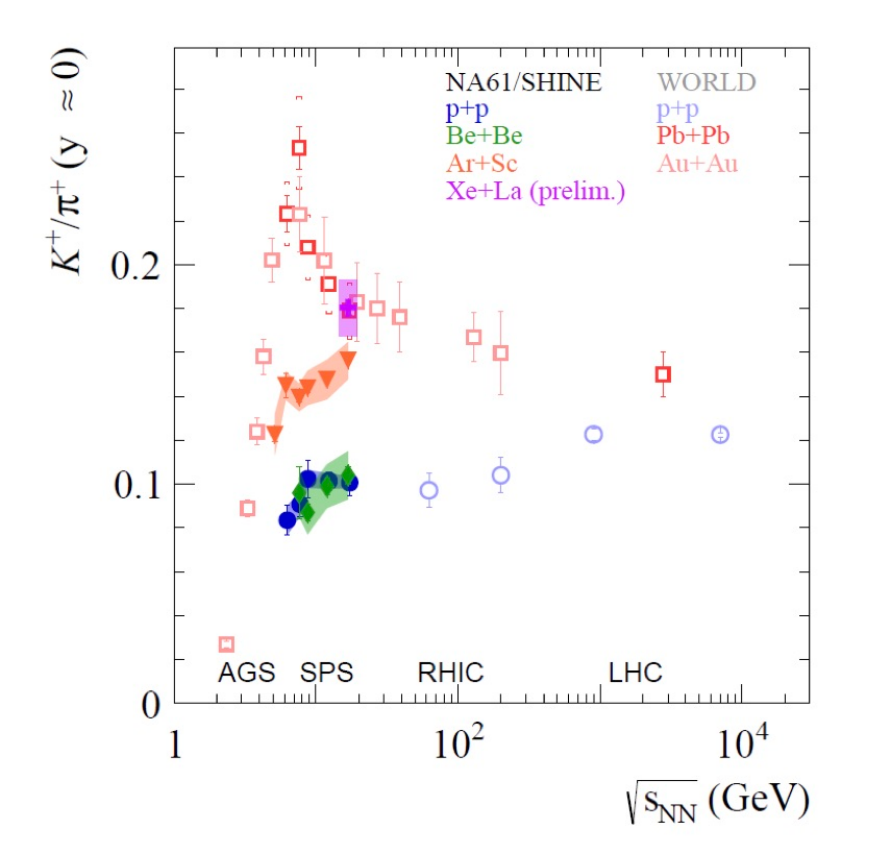}
\end{center}
\caption{
\textit{Left:} Data recorded by NA61/SHINE for its 2D scan programme.  
\textit{Right:} The $K^+$/$\pi^+$ ratio at mid-rapidity as a function of collision energy~\cite{Kowalski:2916893}.  
NA61/SHINE results are presented alongside a compilation of world data.
}
\label{fig:na61_boxplot}      
\end{figure}

Figure~\ref{fig:na61_boxplot}~(\textit{right}) summarises the collision-energy dependence of the positively charged kaon-to-pion ratio, $K^+$/$\pi^+$, at mid-rapidity, as measured by NA61/SHINE~\cite{Kowalski:2916893}. The data include \textit{p+p}, Be+Be, Ar+Sc, Xe+La, and Pb+Pb collisions, alongside a compilation of world data. 

Firstly precise measurements of \textit{p+p} interactions revealed a \textit{break} in the $\sqrt{s_{NN}}$ dependence at around 10~GeV~\cite{NA61SHINE:2019xkb}. This break was initially interpreted as either a transition from resonance- to string-dominated hadron production or a remnant of the onset of deconfinement observed in Pb+Pb collisions. Subsequent results for central Be+Be collisions~\cite{NA61SHINE:2020ggt,NA61SHINE:2020czq}, which surprisingly overlapped with the \textit{p+p} data, clearly supported the former interpretation. This is because the Statistical Model of the Early Stage (SMES) generalised to collisions of low-mass nuclei~\cite{Poberezhnyuk:2015wea} predicted that Be+Be collisions would behave similarly to Pb+Pb collisions, rather than resembling \textit{p+p} interactions.

Thus, it was concluded that the system created in collisions of low-mass nuclei at SPS and below is far from equilibrium. Statistical models cannot approximate its properties. The concepts of the phases of matter, characterised by temperature and chemical potentials, are inapplicable. We had to move beyond the phase diagram of strongly interacting matter to understand the experimental results. This led us to the \textit{diagram of high-energy nuclear collisions}~\cite{Andronov:2022cna}.

Then, results from central Ar+Sc collisions~\cite{NA61SHINE:2021nye,NA61SHINE:2023epu} added another piece to the puzzle. At the top SPS energy, the $K^+$/$\pi^+$ ratio for Ar+Sc collisions matched that of Pb+Pb, but at lower SPS energies, it fell between the \textit{p+p} and Pb+Pb values, showing no horn. The Xe+La ratio~\cite{Kowalski:2916893} at the top SPS energy was consistent with the Pb+Pb and Ar+Sc ratios. Data for lower energies in Xe+La collisions are expected soon.

\begin{figure}[htbp]
\begin{center}
\includegraphics[scale=0.34]{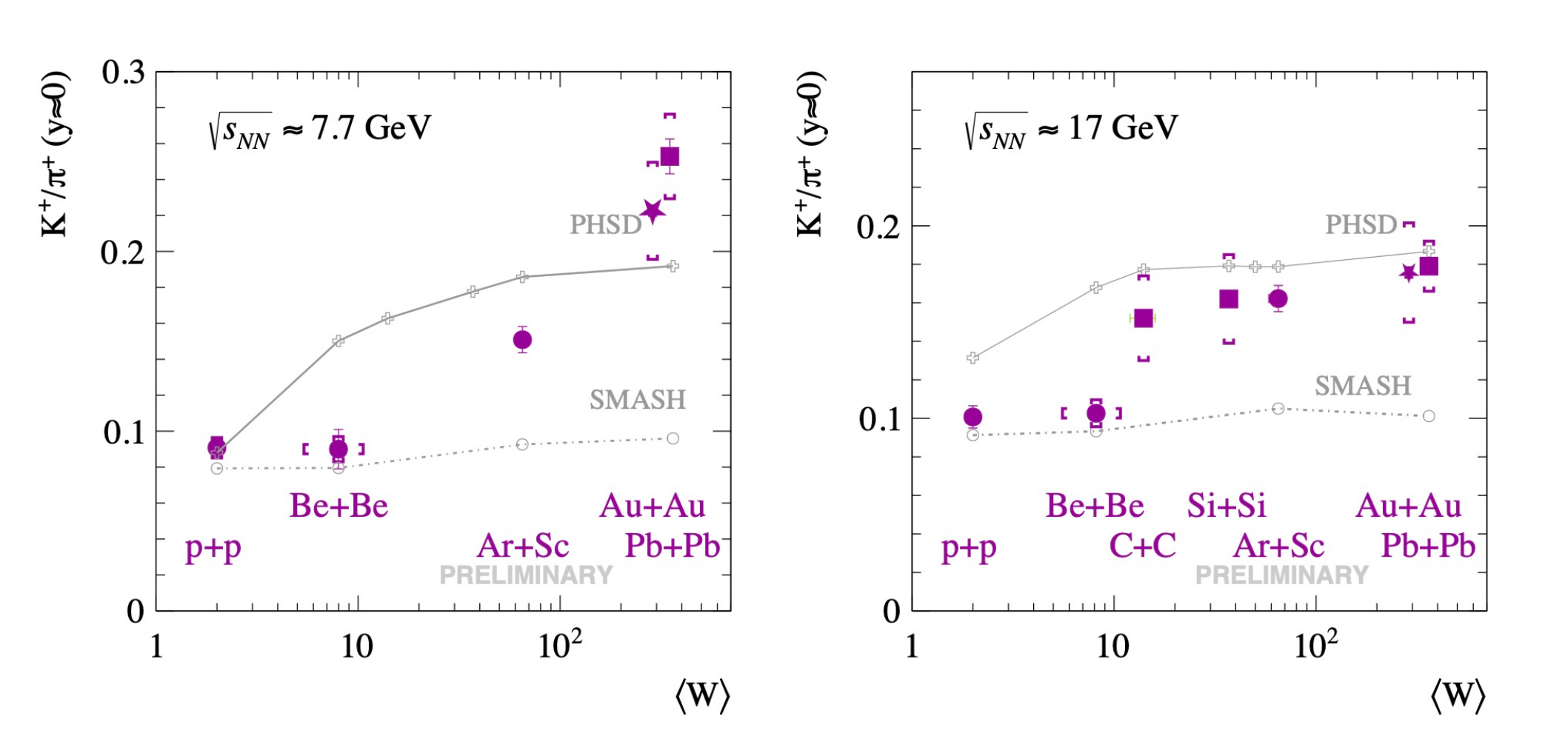}
\end{center}
\caption{
The $K^+$/$\pi^+$ ratio at mid-rapidity measured at $\sqrt{s_{NN}} \approx 7.7$~GeV (\textit{left}) and $\sqrt{s_{NN}} \approx 17$~GeV (\textit{right}) is shown as a function of the mean number of wounded nucleons~\cite{Mackowiak-Pawlowska:2867952}.  
Experimental results are compared with model predictions.
}
\label{fig:na61_jump}      
\end{figure}

The dependence of the $K^+$/$\pi^+$ ratio at the top SPS energy exhibits a rapid change, or \textit{jump}, between collisions of small-mass nuclei (\textit{p+p} and Be+Be) and those of medium- and large-mass nuclei (Ar+Sc, Xe+La, and Pb+Pb). In contrast, at lower energies, the dependence is continuous. Current models fail to describe these observations, as illustrated in Fig.~\ref{fig:na61_jump}, showing the data and the model prediction at 30$A$~(\textit{left}) and
150$A$~GeV/$C$~(\textit{right}) as a function of the mean number of wounded nucleons.

\begin{figure}[htbp]
\begin{center}
\includegraphics[scale=0.34]{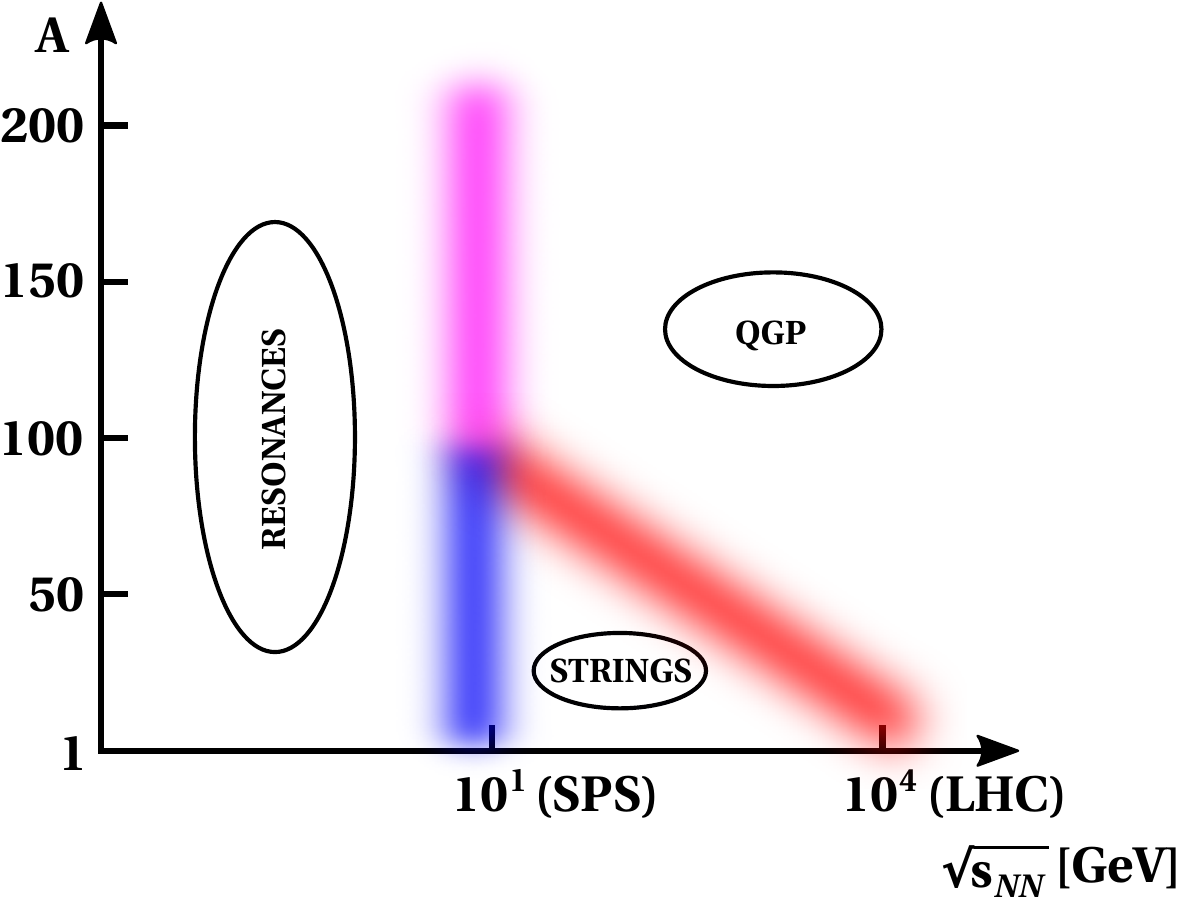}
\includegraphics[scale=0.40]{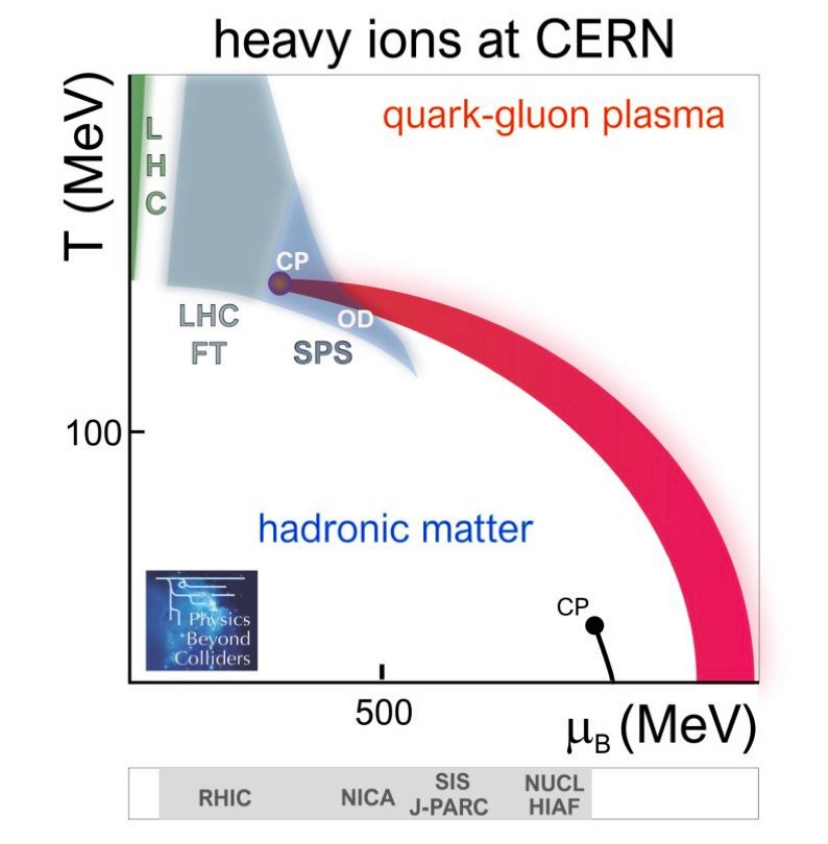}
\end{center}
\caption{
\textit{Left:} Diagram of high-energy nuclear collisions~\cite{Andronov:2022cna}, as suggested by NA61/SHINE and global data.  
\textit{Right:} Hypothetical phase diagram of strongly interacting matter, highlighting the domains accessible for experimental studies with heavy-ion collisions at CERN~\cite{QCDWorkingGroup:2019dyv}.
Up to now, no convincing signal of the critical point ($CP$ on the QGP boundary) has been observed in experiments or demonstrated within QCD. The nuclear liquid-gas critical point ($CP$ at $T \approx$ ~20~MeV) is well established theoretically and experimentally for symmetric nuclear matter.
}
\label{fig:na61_diagrams}      
\end{figure}

As mentioned above, the 2D scan of NA61/SHINE led to the development of a concept of the diagram of high-energy nuclear collisions~\cite{Andronov:2022cna}. 
This diagram depicts the domains dominated by different hadron-production processes in the space of laboratory-controlled parameters: collision energy and the nuclear mass number of the colliding nuclei. A possible interpretation of the results is given in the diagram shown in Fig.~\ref{fig:na61_diagrams}~(\textit{left}).  The lines indicate the changeover between the resonance, string, and QGP domains, with their approximate locations derived from experimental data; for details, see Ref.~\cite{Andronov:2022cna}.

The experimental results suggest that the system created in collisions of high-mass nuclei is close to (local) equilibrium; see, for example, Ref.~\cite{Becattini:2005xt}. In this case, experiments with heavy-ion collisions provide insights into the \textit{phase diagram of strongly interacting matter}. The most widely accepted phase diagram, showing the regions explored by the CERN SPS and LHC experiments, is depicted in Fig.~\ref{fig:na61_diagrams}~(\textit{right}). 

The phase diagram includes the critical point (CP)—the endpoint of the first-order phase transition line, which has properties of a second-order phase transition. 
A significant objective of the NA61/SHINE 2D scan programme has been the search for the CP. Various experimental strategies, inspired by diverse model predictions, have been pursued; however, no signal has been observed to date. CP-exclusion plots have been produced based on the \textit{intermittency} analysis of particle multiplicity fluctuations~\cite{NA61SHINE:2023gez,NA61SHINE:2024ffp}.

During Run~3, NA61/SHINE is collecting high-statistics data on Pb+Pb collisions at the top SPS energy required for systematic measurements of open charm hadrons at the top SPS energy~\cite{Aduszkiewicz:2309890}.  
    A detailed 2D scan, essential for studying the transition from string-dominated to resonance-dominated hadron production~\cite{Mackowiak-Pawlowska:2867952}, is planned following CERN's Long Shutdown~3. The idea~\cite{Gazdzicki:2023niq, Gazdzicki:2025jng} of measuring the correlation between charm and anti-charm hadrons in central Pb+Pb collisions at the top SPS energy is also under consideration~\cite{NA61SHINE:2025aey}.

\subsection{Memories}
\label{subsec:na61_mem}

Actually, I do not remember when and how the idea of a diagram of high-energy nuclear collisions emerged. Scanning the NA61/SHINE repository of public talks, I found the first sketch of it, in my talk presented at the Workshop on the Standard Model and Beyond in Corfu, September 2017~\cite{Corfu_2017}. We called it the ''cross plot''. The current version (Fig.~\ref{fig:na61_diagrams}~(\textit{left})) was triggered by the SHINE Open Seminar~\cite{NA61_SOS} in Zoom on 21 April 2021. Magda (Kuich) presented a talk entitled
"Summary of NA61/SHINE results on the Onset of Fireball" and Edward (Shuryak) followed with the talk "Formation of fireball and clustering near the (hypothetical) critical point". Clearly, under the pressure of new data and ideas, the diagram will continue to evolve, thereby inspiring future experimental measurements.

Twenty years have passed since the initial idea for the NA61/SHINE 2D scan was formulated, as shown in Fig.~\ref{fig:na61_bkg}. Data collection for the scan was completed in 2017, and publication of the main results is expected to conclude within the next few years. Several generations of doctoral and postdoctoral researchers contributed to the results presented in Fig.~\ref{fig:na61_boxplot}~(\textit{right}): Antoni (Aduszkiewicz) and Szymon (Pulawski) worked on \textit{p+p} interactions; Magda (Kuich), Emil (Kaptur), and Szymon on Be+Be collisions; Michel (Naskret), Maciek (Lewicki), and Piotrek (Podlaski) on Ar+Sc collisions; and Sasha (Panova) and Liza (Zherebstova) on Xe+La collisions. Additionally, many contributed to the search for the critical point of strongly interacting matter, with Tobiasz (Czopowicz), Nikos (Davis), Haradhan (Adhikary), and Valeria (Reyna) obtaining the CP exclusion limits.

This period coincided with dramatic global events that also impacted NA61/SHINE: the COVID-19 pandemic, the aggression of Putin's Russia against Ukraine, Trump's chaos and the AI revolution.

\section*{Acknowledgements}
This work was supported by the Polish National Science Centre grant 2018/30/A/ST2/00226.

\bibliographystyle{unsrt}
\bibliography{references}

@article{Kapusta:1979fh,
    author = "Kapusta, Joseph I.",
    title = "{Quantum Chromodynamics at High Temperature}",
    doi = "10.1016/0550-3213(79)90146-9",
    journal = "Nucl. Phys. B",
    volume = "148",
    pages = "461--498",
    year = "1979"
}

@article{Shuryak:1980tp,
    author = "Shuryak, Edward V.",
    title = "{Quantum Chromodynamics and the Theory of Superdense Matter}",
    doi = "10.1016/0370-1573(80)90105-2",
    journal = "Phys. Rept.",
    volume = "61",
    pages = "71--158",
    year = "1980"
}

@techreport{na35_roots,
      author = "NA35",
      title         = "{Study of relativistic nucleus-nucleus reactions induced
                       by $^{16}O$ beams of 9-13 GeV per nucleon at the CERN PS}",
      institution   = "CERN",
      number        = "CERN-PSCC-82-1, PSCC-P-53",
      address       = "Geneva",
      year          = "1982",
      note = {\url{https://cds.cern.ch/record/725513}}
}

@article{Klapisch:1984yfi,
    author = "Klapisch, R.",
    editor = "Ludlam, T. W. and Wegner, H. E.",
    title = "{STATUS OF CERN PROGRAM IN RELATIVISTIC ION PHYSICS}",
    doi = "10.1016/0375-9474(84)90559-1",
    journal = "Nucl. Phys. A",
    volume = "418",
    pages = "347C--352C",
    year = "1984"
}

@article{Schukraft:2015dna,
    author = "Schukraft, Jurgen and Stock, Reinhard",
    title = "{Toward the Limits of Matter: Ultra-relativistic nuclear collisions at CERN}",
    eprint = "1505.06853",
    archivePrefix = "arXiv",
    primaryClass = "nucl-ex",
    doi = "10.1142/9789814644150_0003",
    journal = "Adv. Ser. Direct. High Energy Phys.",
    volume = "23",
    pages = "61--87",
    year = "2015"
}

@article{NA49:1999myq,
    author = "Afanasiev, S. and others",
    collaboration = "NA49",
    title = "{The NA49 large acceptance hadron detector}",
    reportNumber = "CERN-EP-99-001, CERN-EP-99-01, CERN-EP-99-1",
    doi = "10.1016/S0168-9002(99)00239-9",
    journal = "Nucl. Instrum. Meth. A",
    volume = "430",
    pages = "210--244",
    year = "1999"
}

@article{NA61:2014lfx,
    author = "Abgrall, N. and others",
    collaboration = "NA61",
    title = "{NA61/SHINE facility at the CERN SPS: beams and detector system}",
    eprint = "1401.4699",
    archivePrefix = "arXiv",
    primaryClass = "physics.ins-det",
    reportNumber = "CERN-PH-EP-2014-003",
    doi = "10.1088/1748-0221/9/06/P06005",
    journal = "JINST",
    volume = "9",
    pages = "P06005",
    year = "2014"
}

@article{Sandoval:170613,
      author        = "Sandoval, A",
      collaboration = "NA35",
      title         = "{Event simulation in NA35: study of relativistic nucleus -
                       nucleus collisions}",
      reportNumber  = "CERN-EP-86-108",
      journal       = "Nucl. Phys. A",
      volume        = "461",
      pages         = "465-486",
      year          = "1987",
      url           = "https://cds.cern.ch/record/170613",
      doi           = "10.1016/0375-9474(87)90507-0",
}

@article{NA35:1989oqq,
    author = "Gazdzicki, M. and others",
    editor = "Carruthers, P. and Rafelski, Johann",
    collaboration = "NA35",
    title = "{Neutral Strange Particle Production in S S Collisions at 200 GeV/nucleon}",
    doi = "10.1016/0375-9474(89)90613-1",
    journal = "Nucl. Phys. A",
    volume = "498",
    pages = "375C--384C",
    year = "1989"
}

@article{Rafelski:1982pu,
    author = "Rafelski, Johann and Muller, Berndt",
    title = "{Strangeness Production in the Quark - Gluon Plasma}",
    reportNumber = "Print-82-0048 (FRANKFURT)",
    doi = "10.1103/PhysRevLett.48.1066",
    journal = "Phys. Rev. Lett.",
    volume = "48",
    pages = "1066",
    year = "1982",
    note = "[Erratum: Phys.Rev.Lett. 56, 2334 (1986)]"
}

@article{Koch:1986ud,
    author = "Koch, P. and Muller, Berndt and Rafelski, Johann",
    title = "{Strangeness in Relativistic Heavy Ion Collisions}",
    doi = "10.1016/0370-1573(86)90096-7",
    journal = "Phys. Rept.",
    volume = "142",
    pages = "167--262",
    year = "1986"
}

@article{NA35:1988eto,
    author = "Bamberger, A. and others",
    collaboration = "NA35",
    title = "{Probing the Space-time Geometry of Ultrarelativistic Heavy Ion Collisions}",
    doi = "10.1016/0370-2693(88)90561-8",
    journal = "Phys. Lett. B",
    volume = "203",
    pages = "320--326",
    year = "1988"
}

@article{NA35:1990teq,
    author = "Bartke, J. and others",
    collaboration = "NA35",
    title = "{Neutral strange particle production in sulphur sulphur and proton sulphur collisions at 200-GeV/nucleon}",
    doi = "10.1007/BF01554465",
    journal = "Z. Phys. C",
    volume = "48",
    pages = "191--200",
    year = "1990"
}

@techreport{Panagiotou:295042,
      author        = "NA49",
      collaboration = "NA35",
      title         = "{Large acceptance hadron detector for an investigation of
                       Pb-induced reactions at the CERN SPS}",
      institution   = "CERN",
      number        = "CERN-SPSLC-91-31, SPSLC-P-264",
      address       = "Geneva",
      year          = "1991",
      note          = {\url{https://cds.cern.ch/record/295042}}
}

@techreport{Bachler:356588,
      author        = "NA49",
      collaboration = "NA49",
      title         = "{Status and future programme of the NA49 Experiment:
                       addendum-2 to proposal SPSLC/P264}",
      institution   = "CERN",
      number        = "CERN-SPSC-98-4, SPSLC-P-264-Add-2",
      address       = "Geneva",
      year          = "1998",
      note          = {\url{https://cds.cern.ch/record/356588}}
}

@article{NA49:1994hfj,
    author = "Alber, T. and others",
    collaboration = "NA49",
    title = "{Transverse energy production in Pb-208 + Pb collisions at 158-GeV per nucleon}",
    reportNumber = "LBL-37450",
    doi = "10.1103/PhysRevLett.75.3814",
    journal = "Phys. Rev. Lett.",
    volume = "75",
    pages = "3814--3817",
    year = "1995"
}

@article{NA49:1997qey,
    author = "Appelshauser, H. and others",
    collaboration = "NA49",
    title = "{Directed and elliptic flow in 158-GeV / nucleon Pb + Pb collisions}",
    eprint = "nucl-ex/9711001",
    archivePrefix = "arXiv",
    reportNumber = "LBL-41016, LBNL-41016",
    doi = "10.1103/PhysRevLett.80.4136",
    journal = "Phys. Rev. Lett.",
    volume = "80",
    pages = "4136--4140",
    year = "1998"
}

@article{NA49:1997xvz,
    author = "Appelshauser, H. and others",
    collaboration = "NA49",
    title = "{Hadronic expansion dynamics in central Pb + Pb collisions at 158-GeV per nucleon}",
    eprint = "hep-ex/9711024",
    archivePrefix = "arXiv",
    reportNumber = "IKF-HENPG-6-97",
    doi = "10.1007/s100520050168",
    journal = "Eur. Phys. J. C",
    volume = "2",
    pages = "661--670",
    year = "1998"
}

@inproceedings{Roland:1997hs,
    author = "Roland, G.",
    collaboration = "NA49",
    title = "{First results of the NA49 event-by event analysis of Pb + Pb collisions at the SPS}",
    booktitle = "{25th International Workshop on Gross Properties of Nuclei and Nuclear Excitation: QCD Phase Transitions (Hirschegg 97)}",
    pages = "309--318",
    month = "1",
    year = "1997"
}

@article{Gazdzicki:1992ri,
    author = "Gazdzicki, M. and Mrowczynski, S.",
    title = "{A Method to study 'equilibration' in nucleus-nucleus collisions}",
    doi = "10.1007/BF01881715",
    journal = "Z. Phys. C",
    volume = "54",
    pages = "127--132",
    year = "1992"
}

@article{Gorenstein:2011vq,
    author = "Gorenstein, M. I. and Gazdzicki, M.",
    title = "{Strongly Intensive Quantities}",
    eprint = "1101.4865",
    archivePrefix = "arXiv",
    primaryClass = "nucl-th",
    doi = "10.1103/PhysRevC.84.014904",
    journal = "Phys. Rev. C",
    volume = "84",
    pages = "014904",
    year = "2011"
}

@article{Gazdzicki:2011xz,
    author = "Gazdzicki, Marek and Grebieszkow, Katarzyna and Mackowiak, Maja and Mrowczynski, Stanislaw",
    title = "{Identity method to study chemical fluctuations in relativistic heavy-ion collisions}",
    eprint = "1103.2887",
    archivePrefix = "arXiv",
    primaryClass = "nucl-th",
    doi = "10.1103/PhysRevC.83.054907",
    journal = "Phys. Rev. C",
    volume = "83",
    pages = "054907",
    year = "2011"
}

@article{Gorenstein:2011hr,
    author = "Gorenstein, M. I.",
    title = "{Identity Method for Particle Number Fluctuations and Correlations}",
    eprint = "1106.4473",
    archivePrefix = "arXiv",
    primaryClass = "nucl-th",
    doi = "10.1103/PhysRevC.84.024902",
    journal = "Phys. Rev. C",
    volume = "84",
    pages = "024902",
    year = "2011",
    note = "[Erratum: Phys.Rev.C 97, 029903 (2018)]"
}

@article{Rustamov:2012bx,
    author = "Rustamov, A. and Gorenstein, M. I.",
    title = "{Identity Method for Moments of Multiplicity Distribution}",
    eprint = "1204.6632",
    archivePrefix = "arXiv",
    primaryClass = "nucl-th",
    doi = "10.1103/PhysRevC.86.044906",
    journal = "Phys. Rev. C",
    volume = "86",
    pages = "044906",
    year = "2012"
}

@article{Gazdzicki:1996pk,
    author = "Gazdzicki, Marek and Rohrich, Dieter",
    title = "{Strangeness in nuclear collisions}",
    eprint = "hep-ex/9607004",
    archivePrefix = "arXiv",
    reportNumber = "IKF-HENPG-8-95",
    doi = "10.1007/s002880050147",
    journal = "Z. Phys. C",
    volume = "71",
    pages = "55--64",
    year = "1996"
}

@article{Gazdzicki:1995zs,
    author = "Gazdzicki, M. and Roehrich, D.",
    title = "{Pion multiplicity in nuclear collisions}",
    doi = "10.1007/BF01571878",
    journal = "Z. Phys. C",
    volume = "65",
    pages = "215--223",
    year = "1995"
}

@article{Fermi:1950frz,
    author = "Fermi, Enrico",
    title = "{High Energy Nuclear Events}",
    doi = "10.1143/ptp/5.4.570",
    journal = "Prog. Theor. Phys.",
    volume = "5",
    number = "4",
    pages = "570--583",
    year = "1950"
}

@article{Landau:1953wku,
    author = "Landau, Lev Davidovich",
    editor = "ter Haar, D.",
    title = "{On Multiple Production of Particles during Collisions of Fast Particles}",
    doi = "10.1016/b978-0-08-010586-4.50079-1",
    journal = "Izv. Akad. Nauk Ser. Fiz.",
    volume = "15",
    year = "1953"
}

@article{Gazdzicki:1995ze,
    author = "Gazdzicki, M.",
    title = "{Entropy in nuclear collisions}",
    doi = "10.1007/BF01579641",
    journal = "Z. Phys. C",
    volume = "66",
    pages = "659--662",
    year = "1995"
}

@article{Gazdzicki:1998vd,
    author = "Gazdzicki, Marek and Gorenstein, Mark I.",
    title = "{On the early stage of nucleus-nucleus collisions}",
    eprint = "hep-ph/9803462",
    archivePrefix = "arXiv",
    reportNumber = "IKF-HENPG-2-98",
    journal = "Acta Phys. Polon. B",
    volume = "30",
    pages = "2705",
    year = "1999"
}

@article{NA49:2002pzu,
    author = "Afanasiev, S. V. and others",
    collaboration = "NA49",
    title = "{Energy dependence of pion and kaon production in central Pb + Pb collisions}",
    eprint = "nucl-ex/0205002",
    archivePrefix = "arXiv",
    doi = "10.1103/PhysRevC.66.054902",
    journal = "Phys. Rev. C",
    volume = "66",
    pages = "054902",
    year = "2002"
}

@article{NA49:2007stj,
    author = "Alt, C. and others",
    collaboration = "NA49",
    title = "{Pion and kaon production in central Pb + Pb collisions at 20-A and 30-A-GeV: Evidence for the onset of deconfinement}",
    eprint = "0710.0118",
    archivePrefix = "arXiv",
    primaryClass = "nucl-ex",
    doi = "10.1103/PhysRevC.77.024903",
    journal = "Phys. Rev. C",
    volume = "77",
    pages = "024903",
    year = "2008"
}

@article{Gazdzicki:2003dx,
    author = "Gazdzicki, M. and Gorenstein, Mark I. and Grassi, F. and Hama, Yogiro and Kodama, T. and Socolowski, Jr., O.",
    editor = "Navarra, F. S. and Hama, Yogiro",
    title = "{Incident energy dependence of the effective temperature in heavy ion collisions}",
    eprint = "hep-ph/0309192",
    archivePrefix = "arXiv",
    doi = "10.1590/S0103-97332004000200041",
    journal = "Braz. J. Phys.",
    volume = "34",
    pages = "322--325",
    year = "2004"
}

@article{VanHove:1982vk,
    author = "Van Hove, L.",
    editor = "Giovannini, Alberto",
    title = "{Multiplicity Dependence of p(T) Spectrum as a Possible Signal for a Phase Transition in Hadronic Collisions}",
    reportNumber = "CERN-TH-3391",
    doi = "10.1016/0370-2693(82)90617-7",
    journal = "Phys. Lett. B",
    volume = "118",
    pages = "138",
    year = "1982"
}

@article{Gorenstein:2003cu,
    author = "Gorenstein, Mark I. and Gazdzicki, M. and Bugaev, K. A.",
    title = "{Transverse activity of kaons and the deconfinement phase transition in nucleus-nucleus collisions}",
    eprint = "hep-ph/0303041",
    archivePrefix = "arXiv",
    doi = "10.1016/j.physletb.2003.06.043",
    journal = "Phys. Lett. B",
    volume = "567",
    pages = "175--178",
    year = "2003"
}

@techreport{CPOD2024,
      author        = "Gazdzicki, M. and Seyboth, P. and Shuryak, E.",
      title         = "{When Quarks and Gluons Become Free}",
      institution   = "CERN",
      number  = "CERN Courier September 2024",
      address       = "Geneva",
      year          = "2024",
      note          = {\url{https://cerncourier.com/a/when-quarks-and-gluons-become-free/}}
}

@techreport{NA61_history,
      author        = "NA61/SHINE",
      title         = "{History of the Collaboration}",
      institution   = "CERN",
      address       = "Geneva",
      year          = "2024",
      note = {\url{https://shine.web.cern.ch/node/10}}
}

@article{STAR:2010vob,
    author = "Aggarwal, M. M. and others",
    collaboration = "STAR",
    title = "{An Experimental Exploration of the QCD Phase Diagram: The Search for the Critical Point and the Onset of De-confinement}",
    eprint = "1007.2613",
    archivePrefix = "arXiv",
    primaryClass = "nucl-ex",
    note = "arXiv:1007.2613 [nucl-ex]",
    month = "7",
    year = "2010"
}

@article{MPD:2022qhn,
    author = "Abgaryan, V. and others",
    collaboration = "MPD",
    title = "{Status and initial physics performance studies of the MPD experiment at NICA}",
    eprint = "2202.08970",
    archivePrefix = "arXiv",
    primaryClass = "physics.ins-det",
    doi = "10.1140/epja/s10050-022-00750-6",
    journal = "Eur. Phys. J. A",
    volume = "58",
    number = "7",
    pages = "140",
    year = "2022"
}

@article{CBM:2016kpk,
    author = "Ablyazimov, T. and others",
    collaboration = "CBM",
    title = "{Challenges in QCD matter physics --The scientific programme of the Compressed Baryonic Matter experiment at FAIR}",
    eprint = "1607.01487",
    archivePrefix = "arXiv",
    primaryClass = "nucl-ex",
    doi = "10.1140/epja/i2017-12248-y",
    journal = "Eur. Phys. J. A",
    volume = "53",
    number = "3",
    pages = "60",
    year = "2017"
}

@inbook{Gazdzicki:2015oya,
    author = "Gazdzicki, Marek and Gorenstein, Mark I.",
    editor = "Rafelski, Johann",
    title = "{Hagedorn\textquoteright{}s Hadron Mass Spectrum and the Onset of Deconfinement}",
    booktitle = "{Melting Hadrons, Boiling Quarks - From Hagedorn Temperature to Ultra-Relativistic Heavy-Ion Collisions at CERN. With a Tribute to Rolf Hagedorn}",
    eprint = "1502.07684",
    archivePrefix = "arXiv",
    primaryClass = "nucl-th",
    doi = "10.1007/978-3-319-17545-4_11",
    pages = "87--92",
    year = "2016"
}

@inproceedings{Gazdzicki:1994ud,
    author = "Gazdzicki, M.",
    title = "{Pions, baryons and entropy in nuclear collisions}",
    booktitle = "{NATO Advanced Study Workshop on Hot Hadronic Matter: Theory and Experiment}",
    pages = "215--222",
    year = "1994"
}

@techreport{Gazdzicki:685283,
      author        = "NA61/SHINE",
      collaboration = "NA61",
      title         = "{A New Experimental Programme with Nuclei and Proton Beams
                       at the CERN SPS}",
      institution   = "CERN",
      number  = "CERN-SPSC-2003-038, SPSC-EOI-001",
      address       = "Geneva",
      year          = "2003",
      note = {\url{https://cds.cern.ch/record/685283}}
}

@techreport{Gazdzicki:919966,
      author        = "NA61/SHINE",
      collaboration = "NA49-future",
      title         = "{Study of Hadron Production in Collisions of Protons and
                       Nuclei at the CERN SPS}",
      institution   = "CERN",
      number  = "CERN-SPSC-2006-001, SPSC-I-235",
      address       = "Geneva",
      year          = "2006",
      note = {\url{https://cds.cern.ch/record/919966}}
}

@techreport{Gazdzicki:995681,
      author        = "NA61/SHINE",
      collaboration = "NA49-future",
      title         = "{Study of Hadron Production in Hadron-Nucleus and
                       Nucleus-Nucleus Collisions at the CERN SPS}",
      institution   = "CERN",
      number  = "CERN-SPSC-2006-034, SPSC-P-330",
      address       = "Geneva",
      year          = "2006",
      note = {\url{https://cds.cern.ch/record/995681}}
}

@article{NA61SHINE:2011dsu,
    author = "Abgrall, N and others",
    collaboration = "NA61/SHINE",
    title = "{Measurements of Cross Sections and Charged Pion Spectra in Proton-Carbon Interactions at 31 GeV/c}",
    eprint = "1102.0983",
    archivePrefix = "arXiv",
    primaryClass = "hep-ex",
    reportNumber = "CERN-PH-EP-2011-005",
    doi = "10.1103/PhysRevC.84.034604",
    journal = "Phys. Rev. C",
    volume = "84",
    pages = "034604",
    year = "2011"
}

@techreport{vonDoetinchem:2914265,
      author        = "NA61/SHINE",
      collaboration = "NA61/SHINE",
      title         = "{Addendum to the NA61/SHINE Proposal: Request for
                       high-statistics p+p measurements in Run 3}",
      institution   = "CERN",
      number  = "CERN-SPSC-2024-028, SPSC-P-330-ADD-15",
      address       = "Geneva",
      year          = "2024",
      note = {\url{https://cds.cern.ch/record/2914265}}
}

@article{NA61SHINE:2019xkb,
    author = "Aduszkiewicz, A. and others",
    collaboration = "NA61/SHINE",
    title = "{Proton-Proton Interactions and Onset of Deconfinement}",
    eprint = "1912.10871",
    archivePrefix = "arXiv",
    primaryClass = "hep-ex",
    reportNumber = "FERMILAB-PUB-19-664-AD-SCD",
    doi = "10.1103/PhysRevC.102.011901",
    journal = "Phys. Rev. C",
    volume = "102",
    number = "1",
    pages = "011901",
    year = "2020"
}

@article{NA61SHINE:2020ggt,
    author = "Acharya, A. and others",
    collaboration = "NA61/SHINE",
    title = "{Measurements of $\pi^-$ production in $^7$Be+$^9$Be collisions at beam momenta from 19$A$ to 150$A$GeV/$c$ in the NA61/SHINE experiment at the CERN SPS}",
    eprint = "2008.06277",
    archivePrefix = "arXiv",
    primaryClass = "nucl-ex",
    reportNumber = "CERN-EP-2020-150",
    doi = "10.1140/epjc/s10052-020-08514-6",
    journal = "Eur. Phys. J. C",
    volume = "80",
    number = "10",
    pages = "961",
    year = "2020",
    note = "[Erratum: Eur.Phys.J.C 81, 144 (2021)]"
}

@article{NA61SHINE:2020czq,
    author = "Acharya, A. and others",
    collaboration = "NA61/SHINE",
    title = "{Measurements of $\pi^\pm$, $K^\pm$, $p$ and $\bar{p}$ spectra in $^7$Be+$^9$Be collisions at beam momenta from 19$A$ to 150$A$ GeV/$c$ with the NA61/SHINE spectrometer at the CERN SPS}",
    eprint = "2010.01864",
    archivePrefix = "arXiv",
    primaryClass = "hep-ex",
    reportNumber = "CERN-EP-2020-187, CERN-EP-2020-187",
    doi = "10.1140/epjc/s10052-020-08733-x",
    journal = "Eur. Phys. J. C",
    volume = "81",
    number = "1",
    pages = "73",
    year = "2021",
    note = "[Erratum: Eur.Phys.J.C 83, 90 (2023)]"
}

@article{Poberezhnyuk:2015wea,
    author = "Poberezhnyuk, R. V. and Gazdzicki, M. and Gorenstein, M. I.",
    title = "{Statistical Model of the Early Stage of nucleus-nucleus collisions with exact strangeness conservation}",
    eprint = "1502.05650",
    archivePrefix = "arXiv",
    primaryClass = "nucl-th",
    doi = "10.5506/APhysPolB.46.1991",
    journal = "Acta Phys. Polon. B",
    volume = "46",
    number = "10",
    pages = "1991",
    year = "2015"
}

@article{NA61SHINE:2021nye,
    author = "Acharya, A. and others",
    collaboration = "NA61/SHINE",
    title = "{Spectra and mean multiplicities of $\pi ^{-}$ in central${}^{40}$Ar+${}^{45}$Sc collisions at 13A, 19A, 30A, 40A, 75A and 150$A\,\text{ Ge }\text{ V }\!/\!\textit{c}$ beam momenta measured by the NA61/SHINE spectrometer at the CERN SPS}",
    eprint = "2101.08494",
    archivePrefix = "arXiv",
    primaryClass = "hep-ex",
    reportNumber = "CERN-EP-2021-010",
    doi = "10.1140/epjc/s10052-021-09135-3",
    journal = "Eur. Phys. J. C",
    volume = "81",
    number = "5",
    pages = "397",
    year = "2021"
}

@article{NA61SHINE:2023epu,
    author = "Adhikary, H. and others",
    collaboration = "NA61/SHINE",
    title = "{Measurements of $\pi ^\pm $, $K^\pm $, p and $\bar{p}$ spectra in $^{40}\hbox {Ar+}^{45}\hbox {Sc}$ collisions at 13A to 150A~$\text{ Ge }\hspace{-1.00006pt}\text{ V }\!/\!c$}",
    eprint = "2308.16683",
    archivePrefix = "arXiv",
    primaryClass = "nucl-ex",
    reportNumber = "CERN-EP-2023-179, FERMILAB-PUB-23-563-AD",
    doi = "10.1140/epjc/s10052-024-12602-2",
    journal = "Eur. Phys. J. C",
    volume = "84",
    number = "4",
    pages = "416",
    year = "2024"
}

@techreport{Kowalski:2916893,
      author        = "NA61/SHINE",
      collaboration = "NA61/SHINE",
      title         = "{Report from the NA61/SHINE experiment at the CERN SPS}",
      institution   = "CERN",
      number  = "CERN-SPSC-2024-030, SPSC-SR-353",
      address       = "Geneva",
      year          = "2024",
      note = {\url{https://cds.cern.ch/record/2916893}}
}

@article{Andronov:2022cna,
    author = "Andronov, Evgeny and Kuich, Magdalena and Ga\'zdzicki, Marek",
    title = "{Diagram of High-Energy Nuclear Collisions \textdagger{}}",
    eprint = "2205.06726",
    archivePrefix = "arXiv",
    primaryClass = "hep-ph",
    doi = "10.3390/universe9020106",
    journal = "Universe",
    volume = "9",
    number = "2",
    pages = "106",
    year = "2023"
}

@article{Becattini:2005xt,
    author = "Becattini, F. and Manninen, J. and Gazdzicki, M.",
    title = "{Energy and system size dependence of chemical freeze-out in relativistic nuclear collisions}",
    eprint = "hep-ph/0511092",
    archivePrefix = "arXiv",
    doi = "10.1103/PhysRevC.73.044905",
    journal = "Phys. Rev. C",
    volume = "73",
    pages = "044905",
    year = "2006"
}

@article{NA61SHINE:2023gez,
    author = "Adhikary, H. and others",
    collaboration = "NA61/SHINE",
    title = "{Search for the critical point of strongly-interacting matter in $^{40}$Ar ~+~$^{45}$Sc collisions at 150A ~Ge V /c using scaled factorial moments of protons}",
    eprint = "2305.07557",
    archivePrefix = "arXiv",
    primaryClass = "nucl-ex",
    reportNumber = "CERN-EP-2023-082",
    doi = "10.1140/epjc/s10052-023-11942-9",
    journal = "Eur. Phys. J. C",
    volume = "83",
    number = "9",
    pages = "881",
    year = "2023"
}

@article{NA61SHINE:2024ffp,
    author = "Adhikary, H. and others",
    collaboration = "NA61/SHINE",
    title = "{Search for a critical point of strongly-interacting matter in central $^{40}$Ar~+~$^{45}$Sc collisions at 13~A\textendash{}75~A ~GeV/c beam momentum}",
    eprint = "2401.03445",
    archivePrefix = "arXiv",
    primaryClass = "nucl-ex",
    reportNumber = "FERMILAB-PUB-24-0021-AD",
    doi = "10.1140/epjc/s10052-024-13012-0",
    journal = "Eur. Phys. J. C",
    volume = "84",
    number = "7",
    pages = "741",
    year = "2024"
}

@techreport{Aduszkiewicz:2309890,
      author        = "NA61/SHINE",
      collaboration = "NA61/SHINE",
      title         = "{Study of Hadron-Nucleus and Nucleus-Nucleus Collisions at
                       the CERN SPS:   Early Post-LS2 Measurements and Future
                       Plans}",
      institution   = "CERN",
      number  = "CERN-SPSC-2018-008, SPSC-P-330-ADD-10",
      address       = "Geneva",
      year          = "2018",
      note = {\url{https://cds.cern.ch/record/2309890}}
}

@techreport{Mackowiak-Pawlowska:2867952,
      author        = "NA61/SHINE",
      collaboration = "NA61/SHINE",
      title         = "{Addendum to the NA61/SHINE Proposal:  Request for light
                       ions beams in Run 4}",
      institution   = "CERN",
      number  = "CERN-SPSC-2023-022, SPSC-P-330-ADD-14",
      address       = "Geneva",
      year          = "2023",
      note = {\url{https://cds.cern.ch/record/2867952}}
}

@article{Gazdzicki:2023niq,
    author = "Gazdzicki, Marek and Kiko{\l}a, Daniel and Pidhurskyi, Ivan and Tinti, Leonardo",
    title = "{Spatial correlations of charm and anticharm quarks at hadronisation}",
    eprint = "2305.00212",
    archivePrefix = "arXiv",
    primaryClass = "hep-ph",
    doi = "10.1038/s42005-025-02213-y",
    journal = "Commun. Phys.",
    volume = "8",
    number = "1",
    pages = "304",
    year = "2025"
}

@article{Gazdzicki:2025jng,
    author = "Gazdzicki, Marek and Kikola, Daniel and Pidhurskyi, Ivan and Tinti, Leonardo",
    title = "{Apparent teleportation of indistinguishable particles}",
    eprint = "2503.10565",
    archivePrefix = "arXiv",
    primaryClass = "nucl-th",
    month = "3",
    year = "2025"
}

@article{NA61SHINE:2025aey,
    author = "Adhikary, H. and others",
    collaboration = "NA61/SHINE",
    title = "{Proposal from the NA61/SHINE Collaboration for update of European Strategy for Particle Physics}",
    eprint = "2507.08602",
    archivePrefix = "arXiv",
    primaryClass = "nucl-ex",
    reportNumber = "FERMILAB-PUB-25-0462-AD",
    month = "7",
    year = "2025"
}

@article{QCDWorkingGroup:2019dyv,
    author = "Dainese, A. and others",
    collaboration = "QCD Working Group",
    title = "{Physics Beyond Colliders: QCD Working Group Report}",
    eprint = "1901.04482",
    archivePrefix = "arXiv",
    primaryClass = "hep-ex",
    reportNumber = "CERN-PBC-REPORT-2018-008",
    month = "1",
    year = "2019"
}

@techreport{Corfu_2017,
      author        = "Marek Gazdzicki",
      title         = "{Study of Onset of Deconfinement and Search for Critical Point by NA61/SHINE}",
      institution   = "Corfu Summer Institute",
      number  = "",
      address       = "Corfu",
      year          = "2017",
      note = {\url{http://www.physics.ntua.gr/corfu2017/sm.html}}
}

@techreport{NA61_SOS,
      author        = "Kuich, Magda and Shuryak, E",
      title         = "{SHINE Open Seminar}",
      institution   = "CERN",
      number  = "NA61/SHINE website",
      address       = "Geneva",
      year          = "2021",
      note = {\url{https://indico.cern.ch/event/1028768}}
}

@article{NA34:1984occ,
    author = "Gordon, H. and others",
    collaboration = "NA34",
    title = "{STUDY OF HIGH-ENERGY DENSITIES OVER EXTENDED NUCLEAR VOLUMES VIA NUCLEUS NUCLEUS COLLISIONS AT THE SPS.}",
    reportNumber = "CERN-SPSC-84-43",
    month = "6",
    year = "1984"
}

@article{NA36:1992avc,
    author = "Andersen, E. and others",
    collaboration = "NA36",
    title = "{Strangeness production at mid-rapidity in S + Pb collisions at 200-GeV/c per nucleon.}",
    doi = "10.1016/0370-2693(92)91651-O",
    journal = "Phys. Lett. B",
    volume = "294",
    pages = "127--130",
    year = "1992"
}

@article{WA80:1995xza,
    author = "Albrecht, R. and others",
    collaboration = "WA80",
    title = "{Limits on the production of direct photons in 200-A/GeV S-32 + Au collisions}",
    reportNumber = "CERN-PPE-95-186",
    doi = "10.1103/PhysRevLett.76.3506",
    journal = "Phys. Rev. Lett.",
    volume = "76",
    pages = "3506--3509",
    year = "1996"
}

@article{WA85:1991nsm,
    author = "Abatzis, S. and others",
    collaboration = "WA85",
    title = "{Xi-, anti-xi-, Lambda and anti-Lambda production in sulphur - tungsten interactions at 200-GeV/c per nucleon}",
    reportNumber = "CERN-PPE-91-112",
    doi = "10.1016/0370-2693(91)91548-A",
    journal = "Phys. Lett. B",
    volume = "270",
    pages = "123--127",
    year = "1991"
}

@article{WA94:1995szb,
    author = "Abatzis, S and others",
    editor = "Poskanzer, Arthur M. and Harris, J. W. and Schroeder, L. S.",
    collaboration = "WA94",
    title = "{Strange particle production in sulphur-sulphur interactions at 200-GeV/c per nucleon}",
    doi = "10.1016/0375-9474(95)00244-U",
    journal = "Nucl. Phys. A",
    volume = "590",
    pages = "317C--331C",
    year = "1995"
}

@article{NA44:1996xlh,
    author = "Bearden, I. G. and others",
    collaboration = "NA44",
    title = "{Collective expansion in high-energy heavy ion collisions}",
    reportNumber = "CERN-PPE-96-163",
    doi = "10.1103/PhysRevLett.78.2080",
    journal = "Phys. Rev. Lett.",
    volume = "78",
    pages = "2080--2083",
    year = "1997"
}

@article{CERESNA45:1997tgc,
    author = "Agakichiev, G. and others",
    collaboration = "CERES/NA45",
    title = "{Low mass e+ e- pair production in 158/A-GeV Pb - Au collisions at the CERN SPS, its dependence on multiplicity and transverse momentum}",
    eprint = "nucl-ex/9712008",
    archivePrefix = "arXiv",
    doi = "10.1016/S0370-2693(98)00083-5",
    journal = "Phys. Lett. B",
    volume = "422",
    pages = "405--412",
    year = "1998"
}

@article{NA50:2000brc,
    author = "Abreu, M. C. and others",
    collaboration = "NA50",
    title = "{Evidence for deconfinement of quarks and gluons from the J / psi suppression pattern measured in Pb + Pb collisions at the CERN SPS}",
    reportNumber = "CERN-EP-2000-013",
    doi = "10.1016/S0370-2693(00)00237-9",
    journal = "Phys. Lett. B",
    volume = "477",
    pages = "28--36",
    year = "2000"
}

@article{NA52NEWMASS:1996uce,
    author = "Appelquist, G. and others",
    collaboration = "NA52 (NEWMASS)",
    title = "{Strangelet search in Pb Pb interactions at 158-GeV/c per nucleon}",
    reportNumber = "BUHE-96-02",
    doi = "10.1103/PhysRevLett.76.3907",
    journal = "Phys. Rev. Lett.",
    volume = "76",
    pages = "3907--3910",
    year = "1996"
}

@article{WA97:1999uwz,
    author = "Andersen, E. and others",
    collaboration = "WA97",
    title = "{Strangeness enhancement at mid-rapidity in Pb Pb collisions at 158-A-GeV/c}",
    reportNumber = "CERN-EP-99-029, CERN-EP-99-29",
    doi = "10.1016/S0370-2693(99)00140-9",
    journal = "Phys. Lett. B",
    volume = "449",
    pages = "401--406",
    year = "1999"
}

@article{NA57:2006aux,
    author = "Antinori, F. and others",
    collaboration = "NA57",
    title = "{Enhancement of hyperon production at central rapidity in 158-A-GeV/c Pb-Pb collisions}",
    eprint = "nucl-ex/0601021",
    archivePrefix = "arXiv",
    doi = "10.1088/0954-3899/32/4/003",
    journal = "J. Phys. G",
    volume = "32",
    pages = "427--442",
    year = "2006"
}

@article{CERESNA45:2002gnc,
    author = "Adamova, D. and others",
    collaboration = "CERES/NA45",
    title = "{Enhanced production of low mass electron pairs in 40-AGeV Pb - Au collisions at the CERN SPS}",
    eprint = "nucl-ex/0209024",
    archivePrefix = "arXiv",
    doi = "10.1103/PhysRevLett.91.042301",
    journal = "Phys. Rev. Lett.",
    volume = "91",
    pages = "042301",
    year = "2003"
}

@article{NA60:2006ymb,
    author = "Arnaldi, R. and others",
    collaboration = "NA60",
    title = "{First measurement of the rho spectral function in high-energy nuclear collisions}",
    eprint = "nucl-ex/0605007",
    archivePrefix = "arXiv",
    doi = "10.1103/PhysRevLett.96.162302",
    journal = "Phys. Rev. Lett.",
    volume = "96",
    pages = "162302",
    year = "2006"
}

@article{ALICE:2008ngc,
    author = "Aamodt, K. and others",
    collaboration = "ALICE",
    title = "{The ALICE experiment at the CERN LHC}",
    doi = "10.1088/1748-0221/3/08/S08002",
    journal = "JINST",
    volume = "3",
    pages = "S08002",
    year = "2008"
}

@article{ATLAS:2014ipv,
    author = "Aad, Georges and others",
    collaboration = "ATLAS",
    title = "{Measurements of the Nuclear Modification Factor for Jets in Pb+Pb Collisions at $\sqrt{s_{\mathrm{NN}}}=2.76$ TeV with the ATLAS Detector}",
    eprint = "1411.2357",
    archivePrefix = "arXiv",
    primaryClass = "hep-ex",
    reportNumber = "CERN-PH-EP-2014-172",
    doi = "10.1103/PhysRevLett.114.072302",
    journal = "Phys. Rev. Lett.",
    volume = "114",
    number = "7",
    pages = "072302",
    year = "2015"
}

@article{CMS:2015yux,
    author = "Khachatryan, Vardan and others",
    collaboration = "CMS",
    title = "{Evidence for Collective Multiparticle Correlations in p-Pb Collisions}",
    eprint = "1502.05382",
    archivePrefix = "arXiv",
    primaryClass = "nucl-ex",
    reportNumber = "CMS-HIN-14-006, CERN-PH-EP-2015-011",
    doi = "10.1103/PhysRevLett.115.012301",
    journal = "Phys. Rev. Lett.",
    volume = "115",
    number = "1",
    pages = "012301",
    year = "2015"
}

\end{document}